\documentclass[dvips]{arxstspdf}
\usepackage{graphicx}
\usepackage{flushend}
\usepackage{stfloats}

\volume{25}
\issue{4}
\pubyear{2010}
\firstpage{429}
\lastpage{449}
\doi{10.1214/09-STS309}

\makeatletter
\def\Yc{\stackrel{\ldots }{Y}}
\makeatother

\begin{document}
\begin{frontmatter}

\title{Cross-Fertilizing Strategies for Better EM~Mountain Climbing
and DA Field Exploration: A~Graphical Guide Book}
\runtitle{Cross-Fertilizing Strategies for EM Algorithms and DA Samplers}

\begin{aug}
\author[a]{\fnms{David A.} \snm{van Dyk}\ead[label=e1]{dvd@ics.uci.edu}\corref{}} \and
\author[b]{\fnms{Xiao-Li} \snm{Meng}\ead[label=e2]{meng@stat.harvard.edu}}
\runauthor{D. A. van Dyk and X.-L. Meng}

\affiliation{University of California, Irvine and Harvard University}

\address[a]{David A. van Dyk is Professor and Chair, Department of
Statistics, University of California, Irvine, California 92697, USA
\printead{e1}.}
\address[b]{Xiao-Li Meng is Whipple V. N. Jones Professor of Statistics and Chair, Department of Statistics,
Harvard University, Cambridge, Massachusetts
02138, USA
\printead{e2}.}

\end{aug}
%
\begin{abstract}
In recent years, a variety of extensions and refinements have been
developed for data augmentation based model fitting routines. These
developments aim to extend the application, improve the speed and/or
simplify the implementation of data augmentation methods, such as the
deterministic EM algorithm for mode finding and stochastic Gibbs
sampler and other auxiliary-variable based methods for posterior
sampling. In this overview article we graphically illustrate and
compare a number of these extensions, all of which aim to maintain the
simplicity and computation stability of their predecessors. We
particularly emphasize the usefulness of identifying similarities
between the deterministic and stochastic counterparts as we seek more
efficient computational strategies. We also demonstrate the
applicability of data augmentation methods for handling complex models
with highly hierarchical structure, using a high-energy high-resolution
spectral imaging model for data from satellite
telescopes, such as the \textit{Chandra X-ray Observatory}.

\end{abstract}

%
\begin{keyword}
\kwd{AECM}
\kwd{blocking}
\kwd{collapsing}
\kwd{conditional augmentation}
\kwd{ECM}
\kwd{ECME}
\kwd{efficient augmentation}
\kwd{data augmentation}
\kwd{Gibbs Sampling}
\kwd{marginal augmentation}
\kwd{model reduction}
\kwd{NEM}
\kwd{nesting}.
\end{keyword}

\end{frontmatter}
%

\section{Introduction}
\label{sec:intro}

\begin{figure*}[b]

\includegraphics{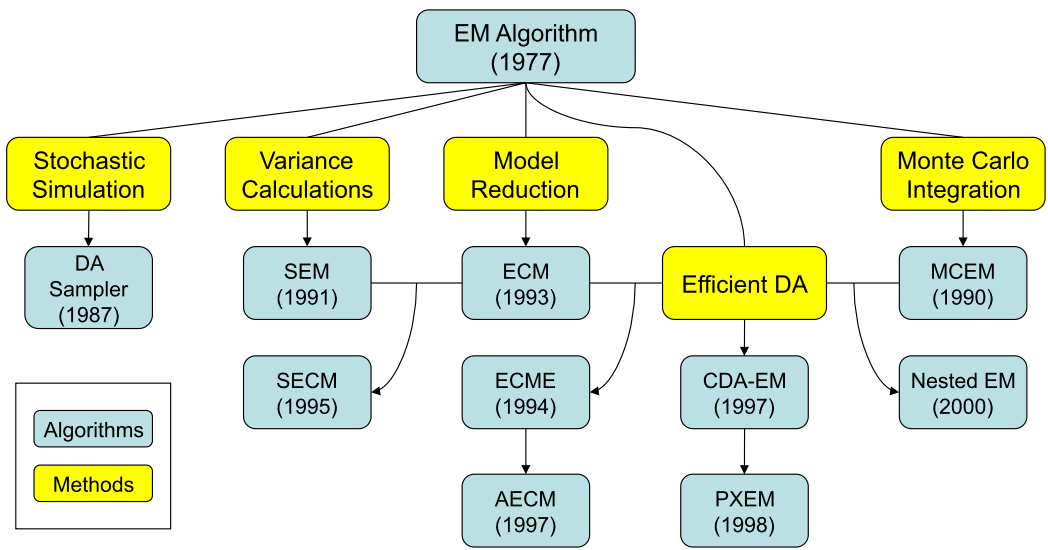}

\caption{A~family tree of algorithms inspired by EM. The~tree
illustrates how various techniques have been combined with the basic
framework of EM to formulate new algorithms. It should be regarded as a
description of the historical inspiration of the various algorithms
rather than as a hierarchy of generalizations and special cases. The~basic \textup{stochastic simulation} EM-type algorithm, known as DA, is
described in Section~\protect\ref{sec:comp:da} and Figure~\protect\ref{fig:em-da}.
\textup{Model reduction} and ECM are described in Section~\protect\ref
{sec:comp:mr} and Figure~\protect\ref{fig:ecm-gibbs}. \textup{Efficient data
augmentation,} including CDA-EM and PXEM, is described in Section~\protect\ref
{sec:eda} and Figures~\protect\ref{fig:condaug} and \protect\ref{fig:pxem}. It is
combined with model reduction to formulate ECME and AECM in
Section~\protect\ref{sec:pcm} and Figure~\protect\ref{fig:aecm}. The~use of \textup{Monte Carlo integration} with and without efficient data augmentation
in MCEM and nested EM is discussed in Section~\protect\ref{sec:eda:nest} and
illustrated in Figures~\protect\ref{fig:partblockgibbs}--\protect\ref{fig:nem}. The~\textup{variance calculations}
of SEM and SECM are developed in Meng and Rubin (\protect\citeyear{mengrubi91}) and van Dyk, Meng and Rubin (\protect\citeyear{vandmengrubi95}), respectively. The~arrows
illustrate the development and combination of techniques that inspired
the generalizations of the EM algorithm.}
\label{fig:emfam}
\end{figure*}

Numerous statistical algorithms involving data\break augmentation have
enjoyed remarkable popularity in the biological, medical, physical,
social, engineering and other sciences. These algorithms include both
deterministic versions such as the Expectation Maximization (EM)
algorithm (\citet{demplairrubi77})
and its many extensions and stochastic versions such as the Data Augmentation (DA)
algorithm (\citeauthor{tannwong87}, \citeyear{tannwong87}), the method of auxiliary
variables (\citeauthor{besagree93}, \citeyear{besagree93}) and other Markov chain Monte
Carlo (MCMC) methods including the Gibbs sampler (\citeauthor{gemagema84},
\citeyear{gemagema84}). The~popularity of these algorithms rests
in their suitability for fitting highly structured models (e.g.,
missing data models, latent variable models, hierarchical models, etc.)
with high dimensional parameters. Such models are themselves growing
ever more popular in modern statistical practice precisely because
complex data generation mechanisms are often naturally defined in terms
of unobserved quantities. This aides inference because the unobserved
quantities often have a direct physical interpretation and are of
scientific interest themselves. From a probabilistic point of view,
complex correlation structures are much more easily described in terms
of unobserved quantities and the conditional independence structures of
hierarchical models. Thus, formulating multi-level models in terms of
unobserved variables enables us to parse complex highly-structured
data. A~primary advantage of algorithms involving data augmentation is
that even in these settings they are relatively easy to implement (as
illustrated in the spectral model of Section~\ref{sec:model}) and
enjoy stable convergence properties (e.g., EM-type algorithms exhibit
monotone convergence in likelihood).

\begin{figure*}

\includegraphics{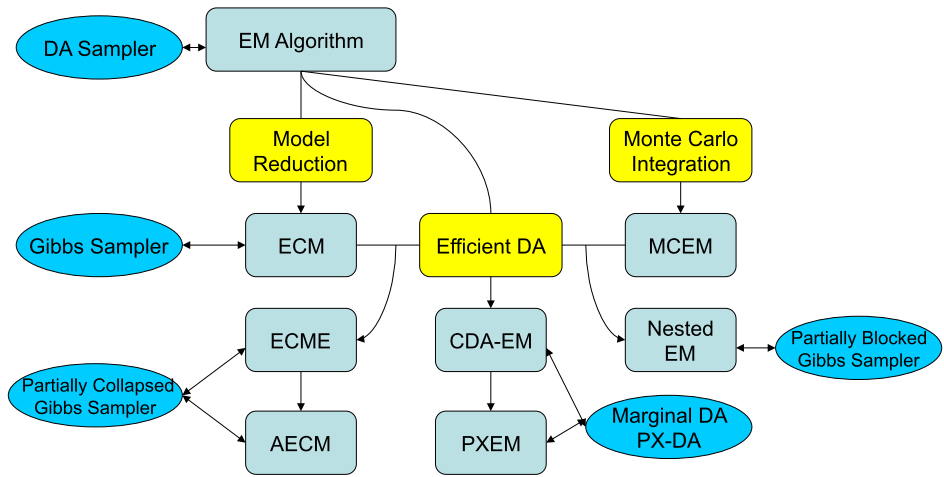}

\caption{The~synergy between EM-type algorithms and their stochastic
counterparts. The~figure shows the cross-fertilization of EM-type
algorithms and DA-type samplers. The~relationships between EM and DA
and between ECM and the Gibbs sample are illustrated in Figures~\protect\ref
{fig:em-da} and \protect\ref{fig:ecm-gibbs}, respectively. Marginal data
augmentation and PX-DA are described in Section~\protect\ref{sec:eda:mar}. The~partially blocked Gibbs sampler that inspired the nested EM algorithm
is illustrated in Figure~\protect\ref{fig:partblockgibbs} and the partially
collapsed Gibbs sampler and its connection with the ECME and AECM
algorithms are discussed in Section~\protect\ref{sec:pcm}.}
\label{fig:dafam}
\end{figure*}

In this paper we review, summarize and compare much of the recent work
on algorithms involving data augmentation, with EM-like algorithms on
the deterministic side and Gibbs-sampler-type MCMC samplers on the
stochastic side. This work is primarily aimed at extending the
applicability of the algorithms and improving their computational
speed. We focus on methods that build on the statistical insight of the
algorithms while maintaining their attractive properties (e.g.,
simplicity and stability), rather than numerical methods that can
sacrifice these properties. We present basic ideas and concepts but
gloss over much of the technical detail, which are documented in the
cited references. To this end, we include a series of schematic graphic
representations of the various algorithms that we hope can clarify and
highlight their relationships, especially in visualizing the
similarities between the deterministic algorithms and their stochastic
counterparts. We begin with two overview schematics. Figure~\ref
{fig:emfam} describes the relationships among the various EM-type
algorithms and Figure~\ref{fig:dafam} describes the synergy between
the deterministic and stochastic algorithms that we discuss in this article.

The~paper is organized into seven additional sections. As a running
example, Section~\ref{sec:model} introduces a model for Poisson
spectral imaging designed to analyze data from the \textit{Chandra X-ray
Observatory} and similar photon counting devices. Section~\ref
{sec:comp} focuses on methods designed to simplify calculation in
complex models, specifically data augmentation and model reduction in
the context of both mode-finding and sampling algorithms. Section~\ref
{sec:rate} reviews general strategies for improving convergence rates
such as blocking and collapsing. These methods are illustrated in
Sections~\ref{sec:eda} and \ref{sec:pcm} in the context of nesting,
conditional augmentation, marginal augmentation, joint augmentation and
partial collapsing. Finally, Section~\ref{sec:examp} applies some of
these methods to the running example and Section~\ref{sec:disc}
concludes with a brief discussion.

\section{A~Poisson Spectral Model}
\label{sec:model}

This section briefly outlines a model for spectral analysis in
astronomy that is designed to summarize high-resolution X-ray and
$\gamma$-ray spectra. The~treatment here is simplified for
illustrational purposes. Details can be found in \citet{vandconnkashsiem01}, \citet{proteta02}, \citet
{hansvand03}, \citet{vandkang04}, \citet{vandetal06} and \citet
{parketal08}. The~spectral model is designed to summarize the
relative frequency of the energy of photons (X-ray or $\gamma$-ray)
arriving at a space-based detector. Because of the digital nature of
the detector, energies are collected as counts in a number of energy
bins (e.g., as many as 4096 on the detectors aboard the \textit{Chandra
X-ray Observatory}). These detectors have much higher resolution than
their predecessors, and thus smaller expected counts per bin.
Independent Poisson distributions are therefore more appropriate to
model the counts than the commonly used Gaussian approximation.

Specifically, we model a spectrum as a mixture of a ``continuum'' term
and an ``emission line.'' The~continuum characterizes the
electromagnetic emission over a broad range of photon energies, while
the emission line can be viewed as an aberration from the continuum in
a narrow range of energies. A~typical spectrum might be composed of
multiple continua and multiple emission lines. For simplicity, we
suppose there is only one of each in the model. In particular, we
parameterize the intensity in bin $j\in\mathcal{J}=\{1,\ldots,J\}$ as
%
\begin{equation}
\quad \lambda_j(\theta) = \delta_j f(\theta^C,E_j) + \nu p_j
(\mu,\sigma^2),\quad     j\in\mathcal{J},
\label{eq:lambda}
\end{equation}
where $\delta_j$ is the known width of bin $j$, $f(\theta^C, E_j)$
represents the continuum term and is a function of the continuum
parameter, $\theta^C$, $E_j$ is the known mean energy in bin $j$, $\nu
$ is the expected photon counts corresponding to the emission line,
$\mu$ and $\sigma$ are the center and scale (or rather ``width'') of
the emission line, and $p_j(\mu,\sigma^2)$, which is a function of
$\mu$ and $\sigma^2$, is the proportion of the emission line counts
that are expected to fall in bin $j$. We typically quantify $p_j(\mu
,\sigma^2)$ via a Gaussian distribution, a $t$ distribution or, in the
case of a very narrow line, a delta function. (These are all standard
astronomical approximations to the distribution of the strictly
positive photon energies of an emission line.) The~collection of
parameters, $\theta^C,(\nu, \mu, \sigma^2)$ and $\theta^A$
(defined below) are together represented by $\theta$. Here we consider
two simple forms of the continuum $f(\theta^C,E_j)$, (1) a log linear
model, for example, the power law $\gamma E_j^{-\beta}$, and (2) a
free (i.e., saturated) model, $f(\theta^C,E_j)=\theta_j^C,$ typically
including a smoothing prior distribution such as a Markov chain for
$\theta_j^C, j\in\mathcal{J}$, for example, $\theta_j^C | \theta_1^C,\ldots,
\theta_{j-1}^C \sim \mathrm{N} (\theta_{j-1}^C, {1 / \omega_j})$
for $j=2,\ldots,J,$ where $\omega= (\omega_2,\ldots,\omega_J)$
is a smoothing parameter and we assume a flat prior distribution for $\theta_1^C$.


Unfortunately, the photon counts are degraded in the observed data. For
example, \textit{instrument response} is a characteristic of the detector
that results in blurring of the photons, that is, a photon that arrives
in bin $j$ has probability $M_{ij}$ of being detected in bin $i\in
\mathcal{I}=\{1,\ldots,I\}$. The~$I\times J$ matrix $\{M_{ij}\}$ is determined by
on-going calibration of the detector and is presumed known. (Because
calibration can be conducted at higher resolution than the binning of
the detector, the instrument response matrix may not be square.)
Another complication is \textit{absorption}, a process by which a
proportion of photons in a given energy bin are absorbed by matter
between the astronomical source and the detector. This results in
stochastic censoring, where the censoring rate varies with energy.
A~similar process occurs in the telescope itself: the detector's
\textit{effective area} depends on the energy of the photons. Finally, the
counts are contaminated by background events. Because of these
degradations, we model the observed counts as independent Poisson
variables with parameters
%
\begin{equation}\label{eq:xi}
\hspace*{12pt}\xi_i(\theta) = \sum_{j=1}^J M_{ij} \lambda_j(\theta) d_j
g(\theta^A, E_j) + \theta_i^B,\quad     i\in\mathcal{I},\hspace*{-12pt}
\end{equation}
where $d_j$ is the (presumed) known effective area of the detector for
energy bin $j$ as a proportion of the total detector area, $g(\theta^A, E_j)$
is the probability that a photon of energy $E_j$ is \textit{not}
absorbed by matter between the source and the detector and $\theta_i^B$
is the Poisson intensity of the background, which is generally
estimated via real-time calibration in space. The~absorption model,
$g(\theta^A,E_j)$, may be a (constrained) log linear model with
$\theta^A$ denoting the model parameter. Note that $\lambda_j(\theta)$
in (\ref{eq:xi}) is given by (\ref{eq:lambda}).

How to construct simple, stable and efficient algorithms for fitting
this model is the running example for the rest of this article.

\section{Statistical Concepts and Computation}
\label{sec:comp}

The~EM algorithm is unique among common numerical optimization routines
in that it is primarily formulated in statistical rather than
mathematical terms. The~missing data setup, the Expectation step and
the complete-data computations of the Maximization step of EM stand in
contrast, for example, to the derivatives and local linearization of
the Newton--Raphson algorithm. Other EM-type optimizers and their
related stochastic samplers extend this in that their motivation and
implementation rely heavily on statistical concepts and insight. In
this section we discuss two such concepts: data augmentation and model
reduction. We show how their effective use of the divide-and-conquer
strategy of reducing a complex problem into an iterated sequence of
simpler ones has led to a rich class of statistical algorithms.

\subsection{Data Augmentation}
\label{sec:comp:da}

\begin{figure*}[b]

\includegraphics{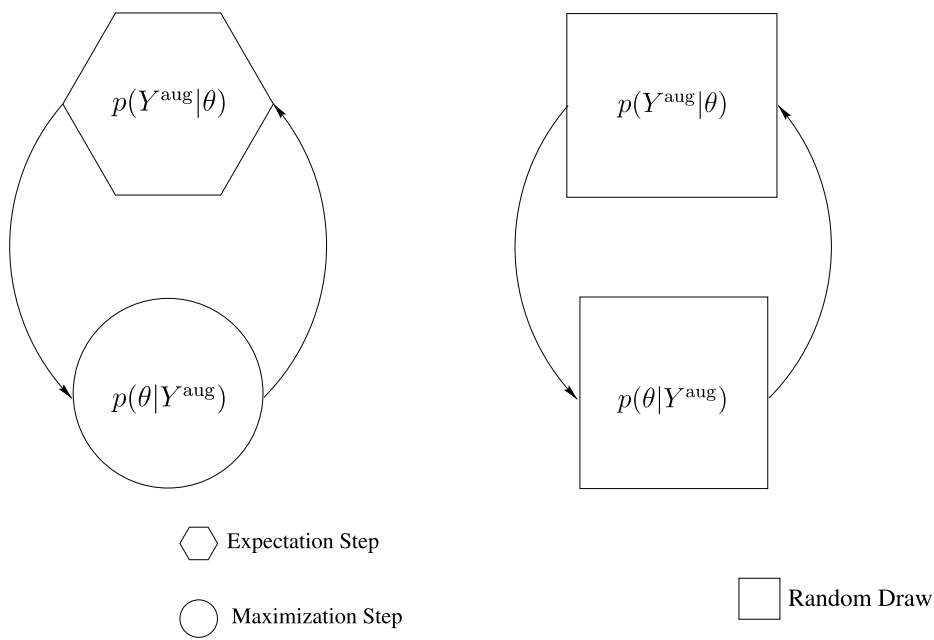}

\caption{The~EM (left panel) and DA (right panel) algorithms. In the
maximization step of EM, we compute $\theta$ to maximize the
conditional expectation of $\log p(\theta|Y^{\mathrm{aug}})$, with the
expectation computed in the expectation step; see Section~\protect\ref{sec:comp:da}.}
\label{fig:em-da}
\end{figure*}

Computational methods based on data augmentation are generally applied
to posterior distributions or likelihood functions. Here we generally
take a Bayesian perspective, but are mindful of the fact that for
computational purposes a likelihood function is equivalent to a
posterior density under a constant prior distribution. Thus, the object
of study can be written as
%
\begin{equation}
p(\theta| Y^{\mathrm{obs}}) = \int p(\theta, \phi|Y^{\mathrm
{obs}}) \mu(d\phi),
\label{eq:margin}
\end{equation}
where $Y^{\mathrm{obs}}$ is the observed data, $\mu$ is a common
measure such as
a Lebesgue or counting measure, $\theta$ is the unobserved quantity of
primary interest, and $\phi$ includes nuisance parameters, latent
variables, missing data or any other unobserved quantity of secondary
interest. Embedding $p(\theta|Y^{\mathrm{obs}})$ into a model on a
larger space such as $p(\theta, \phi|Y^{\mathrm{obs}})$ in this way is
called the method of data augmentation.
This method  can be used to either compute
the mode of $\theta$ under the marginal distribution given in (\ref
{eq:margin}) or to obtain a sample from (\ref{eq:margin}) which in
turn can be used to approximate the posterior mean, variance,
quantiles, etc., via Monte Carlo simulation.

In the spirit of the EM literature, we use a more inclusive notation
$Y^{\mathrm{aug}}$ in place $\phi$, where $Y^{\mathrm{aug}}$ is
called the augmented data
and represents the combination of $Y^{\mathrm{obs}}$ and any latent
variables or
missing data.
The~target posterior distribution can be expressed as
%
\begin{equation}
p(\theta|Y^{\mathrm{obs}}) \propto p(Y^{\mathrm{obs}}|\theta
)p(\theta) ,
\label{eq:post}
\end{equation}
where $p(\theta)$ is a prior distribution and $p(Y^{\mathrm{obs}}|\theta)$
yields a likelihood. In this way, data augmentation methods can be
viewed as embedding (\ref{eq:post}) into a larger \textit{augmented data}
model, via
%
\begin{equation}
\hspace*{8pt} \int_{\mathcal{M}(Y^{\mathrm{aug}})=Y^{\mathrm{obs}}}
p(Y^{\mathrm{aug}}| \theta) \mu(d Y^{\mathrm{aug}}) = p(Y^{\mathrm
{obs}}|\theta) ,\hspace*{-8pt}
\label{eq:defaug}
\end{equation}
where $\mathcal{M}$ is some many-to-one mapping from $Y^{\mathrm{aug}}$
to~$Y^{\mathrm{obs}}$.
Using the factorization
%
\begin{equation}
p(Y^{\mathrm{aug}}| \theta) = p(Y^{\mathrm{aug}}| Y^{\mathrm{obs}},
\theta) p(Y^{\mathrm{obs}}|\theta),
\end{equation}
we recognize that (\ref{eq:defaug}) can be maintained with any choice
of $p(Y^{\mathrm{aug}}| Y^{\mathrm{obs}}, \theta)$, that is, as long
as $p(Y^{\mathrm{aug}}|\theta)$ yields the correct marginal distribution
$p(Y^{\mathrm{obs}}|\theta)$. In
some cases we can use this flexibility to introduce artificial
augmented data purely for computational reasons. Thus, we can choose
$p(Y^{\mathrm{aug}}| Y^{\mathrm{obs}}, \theta)$ in order to optimize
or improve
computational performance rather for statistical modeling, as we shall
discuss in Section~\ref{sec:eda}.

\begin{table*}
\caption{Data augmentation in the spectral model. For all variables,
$j\in\mathcal{J}$, $i\in\mathcal{I}$, and $s\in\mathcal{S}$, where
$\mathcal{J}$ indexes the ideal bins, $\mathcal{I}$ indexes the detector bins, and $\mathcal{S}$
indexes the sources}\label{tbl:specda}
\begin{tabular*}{\textwidth}{@{\extracolsep{\fill}}llcc@{}}
\hline
\textbf{Level} & \multicolumn{1}{c}{\textbf{Variable}} & \textbf{Notation} & \textbf{Range}\\
\hline
1. & The~ideal data: no blurring, binning, background &$ \{\Yc\!{}^C, \Yc\!{}^L\}$& Positive, keV\tabnoteref[b]{b} \\
&  contamination, absorption\tabnoteref[a]{a} or mixing of sources & & \\
2. & The~binned ideal counts &$ \{\ddot Y^C_j, \ddot Y^L_j\}$&Counts\\[3pt]
3. & The~binned ideal counts after absorption & $\{\dot Y^C_j,\dot Y^L_j\}$ & Counts\\[3pt]
4. & The~mixed and binned ideal counts after absorption& $\dot Y_j^+$ &Counts\\[3pt]
5. & The~mixed, binned and blurred ideal counts & $Y_i^+$ & Counts\\
&  after absorption & &\\
6. & The~mixed, binned and blurred ideal counts & $Y^{\mathrm{obs}}_i$& Counts\\
&  after absorption and background contamination, & &\\
&  this is, the observed data & &\\
\hline
\end{tabular*}
\tabnotetext[a]{a}{In the statistical model the effective area of the
instrument is handled in exactly the same way as absorption. Thus, in
this table, absorption includes the effective area of the detector.}
\tabnotetext[b]{b}{The~ideal data are the photon energies measured in
kiloelectron volts (keV).}
\end{table*}

Data augmentation can lead to useful algorithms if the conditional
distributions, $p(Y^{\mathrm{aug}}|Y^{\mathrm{obs}},\theta)$ and
$p(\theta|Y^{\mathrm{aug}})$, are
easy to work with (e.g., to sample, maximize and/or compute
expectations). Thus, a useful choice of an augmented data model
specifies a division of a model into two simpler conditional models
which are typically much easier to analyze.

The~EM algorithm computes a posterior mode using the conditional
distributions via the familiar two-step iteration, consisting of
\begin{enumerate}
\item[E-step:] Compute
\[
Q\bigl(\theta|\theta^{(t)}\bigr)=
\mathrm{E}\bigl[\log p(\theta|Y^{\mathrm{aug}})|Y^{\mathrm{obs}}, \theta
^{(t)}\bigr],
\]
\item[M-step:] Set $\theta^{(t+1)}=\operatorname{argmax}_{\theta} Q(\theta
|\theta^{(t)})$,
\end{enumerate}
where the parenthetical superscript $t$ indexes the iteration. This
iteration is known to increase $p(\theta|Y^{\mathrm{obs}})$ and
converges to a
stationary point of $p(\theta|Y^{\mathrm{obs}})$ that is generally,
but not
always, a (local) mode of $p(\theta|Y^{\mathrm{obs}})$ (\citeauthor{demplairrubi77}, \citeyear{demplairrubi77};
\citeauthor{wu83}, \citeyear{wu83};
\citeauthor{vaid05}, \citeyear{vaid05}). The~two steps of this iteration
give EM its name, that is, the Expectation or E-step and the
Maximization or M-step.

The~Data Augmentation (DA) algorithm of \citet{tannwong87} replaces
the two steps of the EM algorithm with two sampling steps, each samples
one of two full conditional distributions:
\begin{longlist}
\item[Step~1:] $(Y^{\mathrm{aug}})^{(t+1)}\sim p(Y^{\mathrm
{aug}}|Y^{\mathrm{obs}},\theta^{(t)})$,
\item[Step~2:] $\theta^{(t+1)}\sim p(\theta| (Y^{\mathrm{aug}})^{(t+1)})$.
\end{longlist}
This iteration produces a Markov chain, $\{\theta^{(t)},t=1,2,\ldots\}$,
which under mild regularity conditions has the desired stationary distribution,
$p(\theta| Y^{\mathrm{obs}})$ (see \citeauthor{robe96}, \citeyear{robe96};
\citeauthor{tier94}, \citeyear{tier94}, \citeyear{tier96}, for convergence results).
The~EM and DA algorithms are compared in Figure~\ref{fig:em-da}. In all of the figures in this article, conditioning on
$Y^{\mathrm{obs}}$ is suppressed, and hexagons, circles and squares
(or their
elongated versions) represent expectation steps, (conditional)
maximization steps and random draws, respectively.

\subsection{Data Augmentation in the Spectral Model}
\label{sec:comp:da-ex}

Table~\ref{tbl:specda} lists a hierarchy of augmented data structures
used to construct EM and DA~algorithms for fitting the spectral model
described in Section~\ref{sec:model}. In the notation of Table~\ref
{tbl:specda} more dots in the accent above a variable represent greater
degrees of augmentation; variables with fewer dots are (sometimes
stochastic) functions of those with more dots. The~set $\mathcal{S}$ is the
collection of photon sources, here simply $\mathcal{S}=\{C, L\}$, where
$C$ represents the continuum and $L$ the emission line. The~superscript
on ``$Y$'' represents the photon source; a ``$+$'' in the superscript
indicates a mixture of both sources.

\begin{figure*}[b]

\includegraphics{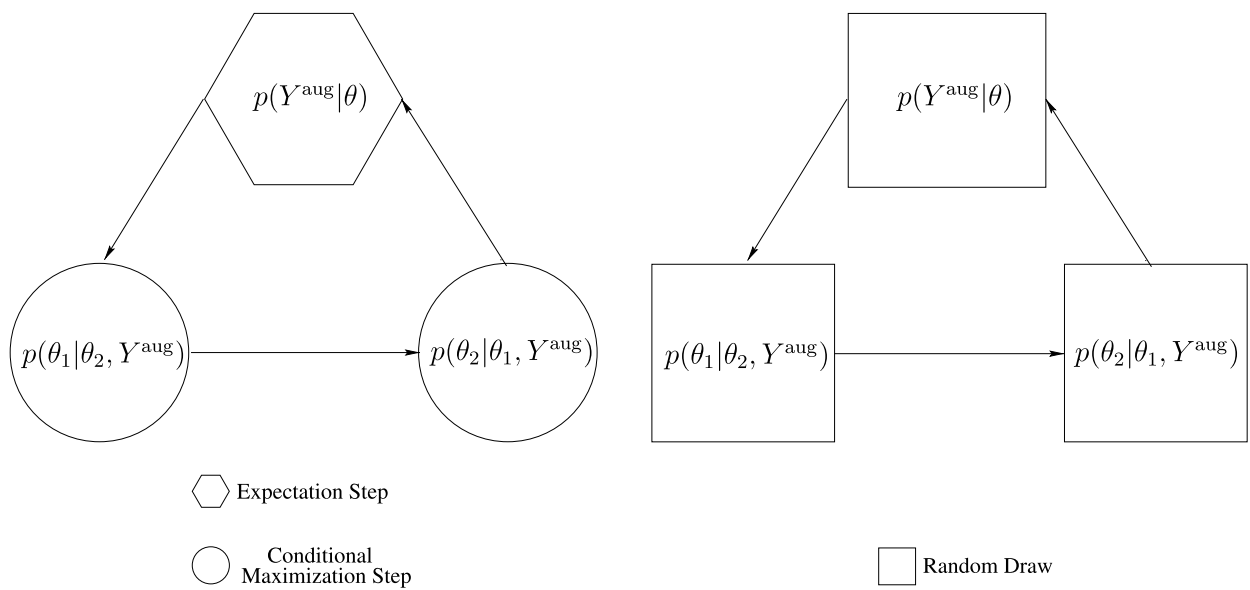}

\caption{The~ECM algorithm and the Gibbs sampler. The~left panel shows
a three-step ECM algorithm composed of an E-step and two CM-steps.
The~corresponding Gibbs sampler is illustrated in the right panel and
is composed of three steps including a data augmentation step. In the
conditional maximization steps of ECM, we compute the component of
$\theta$ to maximize the conditional expectation of the log of the
quantity in the $\bigcirc$, with the expectation computed in the
expectation step; see Section~\protect\ref{sec:comp:mr}.} \label{fig:ecm-gibbs}
\end{figure*}

Reading top-to-bottom in Table~\ref{tbl:specda}, the relationships
among the variables are as follows. The~vectors $\Yc\!{}^C$ and $\Yc\!{}^L$
contain the exact energies of photons attributed to the continuum and
emission line, respectively. Because photon arrivals follow a Poisson
process, the length of both of these vectors are Poisson variables; the
length of $\Yc\!{}^L$ has expectation $\nu$. These energies are binned
and the resulting counts recorded as $\ddot Y^s=(\ddot Y^s_1,\ldots,\ddot Y^s_J)$,
for $s\in\mathcal{S}$. Absorption and the varying effective area
of the instrument cause an energy-varying proportion of these counts to
be lost. In particular,
\begin{eqnarray}
&&\dot Y_j^s| \ddot Y_j^s,\theta\sim\operatorname{Binomial}(\ddot Y_j^s,d_jg(\theta^A,E_j)),\nonumber\\[-8pt]\\[-8pt]
\eqntext{ j\in\mathcal{J},
s\in
\mathcal{S}.}
\end{eqnarray}
For the observer, the continuum and emission line counts are combined,
$\dot Y_j^+ = \dot Y_j^C + \dot Y_j^L$ for each $j$.\vspace*{1pt}
Blurring, due to instrument response, shuffles photons among the bins
and into the observed bin counts via
%
\begin{equation}
Y^+| \dot Y^+, \theta\sim\sum_{j=1}^J \operatorname{Multinomial}
(\dot Y^+_j,M_j),
\end{equation}
where $Y^+=(Y^+_1,\ldots,Y^+_I)$,
$\dot Y^+=(\dot Y^+_1,\ldots,\dot Y^+_J)$, and $M_j$ is the
$j$th column of $M$, $j\in\mathcal{J}$. Because $M$ may not be a square
matrix, the lengths of $Y^+$ and $\dot Y^+$ may differ.
Finally, background contamination leads to the observed bin counts,
%
\begin{equation}
Y^{\mathrm{obs}}_i | Y^+_i,\theta\sim Y^+_i + \operatorname{Poisson}(\theta_i^B),
 \quad   i\in\mathcal{I}.
\end{equation}

This augmented-data construction leads to easy implementation for two
reasons. First, each level of augmented data follows a standard
distribution given $\theta$ and the data in the rows lower in
Table~\ref{tbl:specda}. Reading Table~\ref{tbl:specda} bottom-to-top,
each conditional distribution can be derived using the Bayes theorem.
For illustration, we report the details of just two:
%
\begin{eqnarray}
&&\quad Y^+_i  |  Y^{\mathrm{obs}}_i, \theta\sim\operatorname{Binomial} \biggl(Y^{\mathrm
{obs}}_i,
{{\xi_i(\theta)-\theta_i^B} \over{\xi_i(\theta)}}\biggr),\nonumber\\[-8pt]\\[-8pt]
\eqntext{ i\in
\mathcal{I},}
\end{eqnarray}
where $\xi_i(\theta)$ is defined in (\ref{eq:xi}),
and
%
\begin{equation}
\ddot Y_j^L  |  \dot Y_j^L,\theta\sim\dot Y_j^L + \operatorname{Poisson}
(\eta_{j}),\quad   j\in\mathcal{J},
\end{equation}
where $\eta_{j}=\nu p_j(\mu,\sigma^2)(1-d_jg(\theta^A,E_j))$. The~other necessary conditional distributions can be found in Appendix~B of
\citet{vandconnkashsiem01}. Thus, the E-step of EM and the
corresponding draw of DA are straightforward. Second, given the data in
Table~\ref{tbl:specda}, the posterior distribution of $\theta$ is a
set of independent standard\vspace*{-1pt} distributions. For example, given $\Yc\!{}^L
\buildrel \mathrm{i.i.d.}\over\sim \mathrm{N}(\mu,\sigma^2)$, it is easy to compute
the posterior
distribution of $(\nu,\mu,\sigma^2)$, recalling that the length of
$\Yc\!{}^L$ is a Poisson random variable with mean $\nu$. The~posterior
distributions of the other components of $\theta$ are also standard
and simple to derive. Thus, the M-step of EM and the corresponding draw
of DA are again easy to implement. Incorporating proper prior
information can be accomplished using the appropriate semi-conjugate
prior distributions as described in \citet{vandconnkashsiem01}.

\subsection{Model Reduction}
\label{sec:comp:mr}

Model reduction involves using a set of (typically complete)
conditional distributions in a computation method designed to learn
about the corresponding joint distribution. Reducing the augmented-data
model significantly broadens the applicability of algorithms involved
in data augmentation, while maintaining their stable convergence
properties (e.g.,\break\citeauthor{mengrubi93}, \citeyear{mengrubi93}). In particular, if we
partition $\theta$ into $P$ subvectors, $\theta=(\theta_1,\ldots,\theta_P)$,
reducing the augmented data model involves working with
the set of conditional distributions $p(\theta_1|Y^{\mathrm{aug}},\theta_{-1}),
\break\ldots, p(\theta_P|  Y^{\mathrm{aug}},\theta_{-P})$  in~place of
directly working with $p(\theta|Y^{\mathrm{aug}})$; here  $\theta_{-p}=(\theta_1,
\ldots, \theta_{p-1}, \theta_{p+1},\ldots,\theta_P)$. For
example, the ECM algorithm (\citeauthor{mengrubi93}, \citeyear{mengrubi93}) replaces the
maximization in the M-step of EM with a sequence of $P$ conditional
maximizations or CM-steps of the  form
\begin{enumerate}
\item[CM-step $p$:] Set
$\theta^{(t+{p/P})} = \operatorname{argmax}_\theta Q(\theta| \theta^{(t)})$
\end{enumerate}
subject  to $\theta_{-p}^{(t+{p/P})} = \theta_{-p}^{(t+{(p-1)/P})}$.
The~ECM algorithm is useful when the CM-steps exist in closed form
but the M-step does not. ECM is illustrated with $P=2$ in the left
panel of Figure~\ref{fig:ecm-gibbs}.

The~same strategy can be applied to the DA sampler. By replacing the
draw from $p(\theta|Y^{\mathrm{aug}})$ with a sequence of draws from the
corresponding full conditional distributions, the sampler becomes a
$(P+1)$-step Gibbs sampler. This sampler is illustrated in the right
panel of Figure~\ref{fig:ecm-gibbs}. In the context of sampling, we
can also reduce $p(Y^{\mathrm{aug}}| \theta)$ into a set of conditional
distributions. Partitioning the expectation step, however, has proven
much more illusive. One strategy involves using the law of iterated
expectations in the computation of the E-step and results in the Nested
EM algorithm; see Section~\ref{sec:eda:nest}.\vadjust{\goodbreak}

Rather than using a partition of $\theta$, a more general model
reduction scheme updates $\theta$ by conditioning on a sequence of
functions of $\theta$. It is only required that the functions allow
movement anywhere in the parameter space, that is, the functions are
``space-filling'' as described by \citet{mengrubi93}. Again, the
same strategy can be used in sampling algorithms, such as the Bayesian
IPF sampler used to fit constrained models on contingency tables (\citeauthor{scha97}, \citeyear{scha97};
\citeauthor{gelmcarlsterubi03}, \citeyear{gelmcarlsterubi03}). Recent work by \citet{yumeng09}
further explores the use of this strategy to improve MCMC algorithms by
employing a sequence of sufficient and auxiliary data augmentation
schemes that are space filling.

\subsection{Model Reduction in the Spectral Model}
\label{sec:comp:mr-ex}

To illustrate model reduction in an augmented data model, we consider
the second form of the continuum model, namely, the free model
$f(\theta^C,E_j)=\theta_j^C$ with a Markov-chain-type smoothing prior\vspace*{1pt}
$\theta_j^C | \theta_1^C,\break \ldots,\theta_{j-1}^C \sim \mathrm{N} (\theta_{j-1}^C,
{1 / \omega_j})$  for $j=2,\ldots,J,$ where\vspace*{1pt}
$\omega= (\omega_2,\ldots,\omega_J)$ is a smoothing parameter and
we assume a flat prior for $\theta_1^C$. For simplicity, we assume
there is no emission line and that $\delta_j = g(\theta^A,E_j)=1$ for
each~$j$, that is, the bins are of the same size and that there is no
absorption. In this case, we use only rows~\mbox{4--6} of Table~\ref
{tbl:specda} in our data augmentation scheme to derive
\begin{eqnarray}\label{eq:Qfree}
Q\bigl(\theta|\theta^{(t)}\bigr)
&=& \sum_{j=1}^J \bigl[\mathrm{E}\bigl(\dot Y_j^C|Y^{\mathrm{obs}},\theta^{(t)}\bigr)
\log\theta_j^C - \theta_j^C \bigr] \nonumber\\[-8pt]\\[-8pt]
&&{}- {1 \over2} \sum_{j=2}^J \omega_j
(\theta_j^C - \theta^C_{j-1})^2 .\nonumber
\end{eqnarray}
Once we have computed the expectation in (\ref{eq:Qfree}), we need
only optimize $Q(\theta|\theta^{(t)})$ as a function of $\theta$.
Unfortunately, this optimization cannot be done analytically when some
$\omega_j>0$. However, the partial derivative of $Q(\theta|\theta^{(t)})$
with respect to $\theta_j^C$ is a quadratic function\vspace*{-1pt} of
$\theta_j^C$ if we fix $\theta_{-j}^C$. Thus, as is discussed by\vspace*{1pt}
\citet{fesshero95} and is improved in Section~\ref{sec:examp}, we
can construct an ECM algorithm with $J$ CM-steps of the form
\begin{eqnarray*}
&&(\theta_j^C)^{(t+1)} \\
&&\quad =\max\biggl\{0,{1 \over{A_j}}\Bigl(B_j + \sqrt{B_j^2 +
A_j \mathrm{E}\bigl(\dot Y_j^C | Y^{\mathrm{obs}},\theta^{(t)}\bigr)}\Bigr) \biggr\},
\end{eqnarray*}
where
\begin{eqnarray*}
A_j &=& \omega_j + \omega_{j+1}
 \quad   \mbox{and}\\
B_j&=& - \bigl( 1 - \omega_j(\theta_{j-1}^C)^{(t+1)}- \omega_{j+1}(\theta
_{j+1}^C)^{(t)}\bigr) / 2 .
\end{eqnarray*}

\section{Improving Rates of Convergence}
\label{sec:rate}

EM-type algorithms and their stochastic counterparts have seen many
applications largely because of their computational stability and
simple implementation. Nonetheless, these methods are legitimately
criticized for their slow convergence in some settings. Strong
posterior correlations among the components updated in each step lead
to full conditional distributions that are far less variable than the
corresponding marginal distributions. This in turn leads to smaller
step sizes and slower progress toward the mode or toward the stationary
distribution. Much work has been focused on developing algorithms with
improved rates of convergence that continue to enjoy the simplicity and
stability that makes data augmentation so useful in practice. As we
shall see with both data augmentation and model reduction, \textit{less is
better} if one hopes for speed, while more is often better if one hopes
for simplicity. In this section we discuss the sometimes conflicting
strategies for improving the computational performance of methods based
on data augmentation.

\subsection{The~EM and DA Rates of Convergence}
\label{sec:rate:em-da}

Before we can develop criteria for speeding up data augmentation
methods, we need mathematical measures of their rates of convergence.
For EM, such a measure is given by $\rho_{\mathrm{EM}}$, the spectral
radius of the so-called matrix fraction of missing information (\citeauthor{demplairrubi77}, \citeyear{demplairrubi77}),
%
\begin{equation}
I-I^{\mathrm{obs}}[I^{\mathrm{aug}}(Y^{\mathrm{aug}})]^{-1},
\label{eq:emrate}
\end{equation}
where $I$ is an identity matrix, $I^{\mathrm{obs}}$ is the observed Fisher
information matrix, and $I^{\mathrm{aug}}(Y^{\mathrm{aug}}) =\break
-{\partial^2}Q(\theta| \theta^{\ast})/{(\partial\theta\,\partial\theta)}
|_{\theta=\theta^{\ast}}$ with $\theta^{\ast}$ the posterior
mode; our notation for $I^{\mathrm{aug}}$ emphasizes that both
$Q(\theta| \theta^\prime)$ and the augmented-data information matrix depend on the
choice of augmented data model. Here we use the traditional terms (e.g.,
Fisher information) of the EM literature, which primarily focus on
likelihood calculation, even though we are dealing with the more
general posterior computation. In particular, $I^{\mathrm{obs}}$ is
the negative
of the second derivative of the log posterior density evaluated at the
posterior mode.

We call $\rho_{\mathrm{EM}}$ the global rate of convergence and
$I-I^{\mathrm{obs}}(I^{\mathrm{aug}})^{-1}$ the matrix rate of convergence of the EM
algorithm. More
general formulations of the rate of convergence for ECM and other
EM-type algorithms are given by \citeauthor{mengrubi93} (\citeyear{mengrubi93},
\citeyear{mengrubi94}), \citet{meng94b}, \citet{mengvand97} and \citet{vand00nem}.
For the EM algorithm, our goal is to minimize $\rho_{\mathrm{EM}}$ as a
function of the data augmentation scheme.

For the DA algorithm, the
geometric rate of convergence (\citeauthor{amit91}, \citeyear{amit91}) is
%
\begin{equation}
\quad 1-\inf_{h:\operatorname{Var}(h(\theta)|Y^{\mathrm{obs}})=1} \mathrm
{E}[\operatorname{Var}(h(\theta)|Y^{\mathrm{aug}}
)|Y^{\mathrm{obs}}].
\label{eq:darate}
\end{equation}
Although this quantity and the maximum lag one autocorrelation (\citeauthor{liu94a}, \citeyear{liu94a}) are valuable for theoretical calculations, they are generally
difficult to work with analytically in particular models. The EM-approximation of \citet{vandmeng01art} is essentially based on a
Gaussian approximation to the posterior distribution and simply
replaces these quantities by $\rho_{\mathrm{EM}}$. Van Dyk and Meng (\citeyear{vandmeng01art})
illustrate that this approximate EM criterion can lead to substantial
improvements in DA samplers. Thus, one of our basic strategies is to
focus on methods that reduce $\rho_{\mathrm{EM}}$ with an understanding
that such methods are useful in formulating efficient data augmentation
schemes for both deterministic and sto\-chastic algorithms.

\subsection{Blocking and Collapsing}
\label{sec:rate:bnc}

As the formulations of the matrix rates of convergence for more complex
EM-type algorithms in the above cited articles illustrate, analysis of
convergence is significantly more complex with multi-step algorithms.
In the analysis of DA and Gibbs samplers, the spectral radius and the
norm of the forward operator are useful measures of the convergence
behavior of a Markov chain (\citet{liuwongkong94};
\citeauthor{liu01}, \citeyear{liu01}). Based on
these measures, \citet{liuwongkong94} introduced two strategies
that have emerged as important general techniques for improving the
behavior of Gibbs-type samplers.

To illustrate these techniques, consider a $P$-step sampler that
simulates each component of $\theta= (\theta_1, \ldots,\break \theta_P)$
in turn conditioning on the most recently sampled values of the other
$P-1$ components of $\theta$. The~first strategy, known as \textit{blocking}, involves combining two or more draws into a single draw. For
example, the last two steps could be combined into a single draw of
$(\theta_{P-1}, \theta_P)$ given the other $P-2$ components of
$\theta$. \textit{Collapsing}, on the other hand, involves the
construction of a sampler on a subspace of the original sampler. For
example, we might compute the marginal distribution of $\theta_{-P}$
by integrating out $\theta_P$ and construct a $(P-1)$-step  sampler
using the full conditional distributions of the first $P-1$ components
of the original partition of $\theta$. Each of these components is
updated conditioning on the most recently sampled values of the other
$P-2$ components of $\theta_{-P}$ to construct a Markov chain with
stationary distribution equal to the marginal distribution of $\theta_{-P}$.

\citet{liu01} shows that both of these strategies are expected to
improve the convergence behavior of the original $P$-step sampler in
that they reduce the norm of its forward operator. (For Gibbs samplers
with more than two steps, the norm may not be equal to the rate of
convergence of the Markov chain.) He also showed that collapsing
reduces the norm by at least as much as blocking. Thus, good general
advice is to collapse whenever possible, and to block if you can when
collapsing is not possible. Liu's technical results apply only when
blocking is applied to the last steps of each iteration of a Gibbs
sampler and/or when the subparameter sampled in the last step is
collapsed out of the sampler, as we discussed for illustration in the
previous paragraph. Nonetheless, experience shows that both strategies
are more generally useful and should be implemented whenever feasible.

Analogous advice applies to EM-type algorithms. In the comparison of
the EM and ECM algorithms, blocking suggests that fewer CM-steps should
be preferred and that the ECM algorithm is expected to converge more
slowly than the corresponding EM algorithm. While this is good general
advice, it does not always hold mathematically; \citet{meng94b} gives
a simple example in which ECM outperforms EM. We emphasize that the
motivation of ECM, however, is not faster convergence but easier
implementation. We generally consider ECM when the M-step of EM is not
tractable and, thus, the EM algorithm itself is not feasible.

Collapsing is also a useful strategy in the context of EM-type
algorithms. The~next section is devoted to methods that aim to reduce
the information in $Y^{\mathrm{aug}}$ and thus effectively collapse a
portion of
$Y^{\mathrm{aug}}$ out of the iteration. Section~\ref{sec:pcm} describes
intermediate strategies that allow partial collapse when full collapse
is not possible, as in the ECME and AECM algorithms.

In the context of EM, we can sometimes also collapse $\theta$ via a
profile loglikelihood. Suppose that $\theta=(\theta_1, \theta_2)$
and that we are able to compute the profile likelihood $\tilde\ell
(\theta_1; Y^{\mathrm{obs}}) = \ell(\theta_1, \hat\theta_2(\theta_1, Y^{\mathrm{obs}})
| Y^{\mathrm{obs}}),$ where $\hat\theta_2(\theta_1, Y^{\mathrm{obs}})$ is the
maximizer of $\ell(\theta_1,\theta_2 | Y^{\mathrm{obs}})$ when
$\theta_1$ is fixed. There are two ways to construct an EM algorithm in this
situation. The~first way is to construct a data augmentation,
$Y^{\mathrm{aug}}$,
to implement EM for the full parameter $\theta=(\theta_1, \theta_2)$
via the full augmented data loglikelihood,
$\ell(\theta_1,\theta_2 | Y^{\mathrm{aug}})$. That is, we do not take advantage of the potential
computational gain of using the profile likelihood. The~second way is
to construct a data augmentation, $\tilde Y^{\mathrm{aug}}$, to
augment the
profile likelihood $\tilde\ell(\theta_1; Y^{\mathrm{obs}})$ and
then implement
the EM algorithm for the subparameter $\theta_1$ only. Note that here
we use the notation $\tilde\ell(\theta_1; Y^{\mathrm{obs}})$ rather than
$\tilde\ell(\theta_1| Y^{\mathrm{obs}})$ to emphasize that
$\tilde\ell(\theta_1; Y^{\mathrm{obs}})$ may not necessarily be a proper
loglikelihood in
the sense of being derived from a log density or probability of
$Y^{\mathrm{obs}}$. We can nonetheless use EM, because it is possible
to construct an EM algorithm for maximizing any objective function
$D(\theta;Y^{\mathrm{obs}})$ as
long as we can find an augmented objective function
$D(\theta; Y^{\mathrm{aug}})$ such that
$\exp\{D(\theta; Y^{\mathrm{aug}}) -D(\theta; Y^{\mathrm{obs}})\}$
is a proper conditional density function of $Y^{\mathrm{aug}}$ given
$\theta$ and $Y^{\mathrm{obs}}$; see the rejoinder of \citet{mengvand97} for
more discussion on this flexibility of EM. Therefore, it is possible to
use EM for the profile likelihood by treating $\tilde\ell(\theta_1; Y^{\mathrm{obs}})$
as an objective function. This collapsing through profiling
has not been generally recognized, but can significantly improve the
speed, when compared to the first way of directly applying the EM
algorithm to the full likelihood. See \citet{meng97} for more
discussion and an example involving a zero inflated Poisson model.

\section{Efficient Data Augmentation}
\label{sec:eda}

Inherent in the definition of the augmented data model is a choice:
There are infinitely many augmented data models satisfying (\ref{eq:defaug}).
In this section we discuss various criteria for this
choice that result in efficient algorithms. By ``efficient data
augmentation'' we mean using augmentation schemes that improve \textit{speed},
while maintaining \textit{stability and simplicity}. Here we
discuss techniques that are able to achieve all three criterion: They
reduce the augmented data in the construction of the algorithm to
improve speed while maintaining stability and simplicity.

The~basic idea is similar to collapsing in the Gibbs sampler. Suppose
that an EM algorithm or a data augmentation sampler can be constructed
with a baseline data augmentation scheme that we denote
${\tilde Y^{\mathrm{aug}}}$. Further suppose that
${\tilde Y^{\mathrm{aug}}}= Y^{\mathrm{aug}}_1 \cup Y^{\mathrm{aug}}_2$,
whe\-re both $Y^{\mathrm{aug}}_1$ and
$Y^{\mathrm{aug}}_2$ are legitimate data augmentation schemes in that
they both
contain $Y^{\mathrm{obs}}$. It is easy to show that
$I^{\mathrm{aug}}({\tilde Y^{\mathrm{aug}}}) \geq I^{\mathrm{aug}}
(Y^{\mathrm{aug}}_1)$ [i.e., that
$I^{\mathrm{aug}}({\tilde Y^{\mathrm{aug}}}) - I^{\mathrm{aug}}(Y^{\mathrm{aug}}_1)$
is semi-positive definite] and that
$\mathrm{E}[\operatorname{Var}(h(\theta)|{\tilde Y^{\mathrm{aug}}})|Y^{\mathrm{obs}}]
\leq\mathrm{E}[\operatorname{Var}(h(\theta)|\break Y^{\mathrm{aug}}_1)|Y^{\mathrm{obs}}]$, where $h(\cdot)$ is any
real-valued function, the first expression being an asymptotic variant
of the second (\citeauthor{mengvand99}, \citeyear{mengvand99}). Thus, by (\ref{eq:emrate}) and
(\ref{eq:darate}), construction of an alternate algorithm using only
$Y^{\mathrm{aug}}_1$ as the augmented data results in faster
convergence. This
strategy effectively collapses $\tilde Y^{\mathrm{aug}}\setminus Y^{\mathrm{aug}}_1$
out of the algorithm. We will discuss direct applications of this idea when we
discuss the nesting strategy in Section~\ref{sec:eda:nest}. Less
direct applications are the topic of Sections~\ref{sec:eda:cond}--\ref
{sec:eda:joint}. The~methods described in these sections do not
directly decompose ${\tilde Y^{\mathrm{aug}}}$ into two components but
still aim to either
reduce $I^{\mathrm{aug}}({\tilde Y^{\mathrm{aug}}})$ or to increase
$\mathrm{E}[\operatorname{Var}(h(\theta)|{\tilde Y^{\mathrm{aug}}})|Y^{\mathrm{obs}}]$.

\subsection{Conditional Augmentation}
\label{sec:eda:cond}
The~methods of conditional, marginal and joint augmentation all take
advantage of the flexibility in (\ref{eq:defaug}) to introduce less
informative augmented data in order to construct a more efficient
algorithm. To search for a good augmented data model using any of the
three methods, we begin by parameterizing the augmented data model
using a \textit{working parameter}. We define a working parameter to be a
parameter in the augmented data model that is not identifiable under
the observed data model, $p(Y^{\mathrm{obs}}|\theta)$. In particular, we
generalize (\ref{eq:defaug}) via
\begin{eqnarray}\label{eq:defwkpar}
&&\int_{\mathcal{M}(Y^{\mathrm{aug}})=Y^{\mathrm{obs}}} p(Y^{\mathrm
{aug}}|\theta, \alpha) \mu(d Y^{\mathrm{aug}})\nonumber\\[-8pt]\\[-8pt]
&&\quad = p(Y^{\mathrm{obs}}|\theta)\nonumber
\end{eqnarray}
for all $\alpha$ in some class $\mathcal{A}$. Notice that the right-hand
side of (\ref{eq:defwkpar}) does not depend on the working parameter.
An effective method of introducing $\alpha$ is to let
$Y^{\mathrm{aug}}=\mathcal{D}_{\alpha, \theta}({\tilde Y^{\mathrm{aug}}})$,
where $\mathcal{D}_{\alpha, \theta}$ is a one-to-one mapping for
any $\theta$ and $\alpha\in\mathcal{A}$ and ${\tilde Y^{\mathrm{aug}}}$ is the baseline
augmented data. Typically ${\tilde Y^{\mathrm{aug}}}$
is the standard augmented data used to construct EM-type algorithms
or samplers for fitting a particular model. In the context of the EM
algorithm, we can compute the scalar rate of convergence,
$\rho_{\mathrm{EM}} (\alpha)$, for each $\alpha$. Conditional augmentation simply
optimizes $\rho_{\mathrm{EM}}(\alpha)$ as function of $\alpha$ and then
conditions on the optimal value of $\alpha$ throughout the iteration.
\citet{mengvand97} call an EM algorithm constructed with the
resulting optimal data augmentation scheme an \textit{efficient data
augmentation EM algorithm}. For clarity, we refer to it here as a
\textit{conditional data augmentation EM algorithm} or CDA-EM. Although this
choice of augmented data model is based on the EM rate of convergence,
the same model can be used to construct data augmentation samplers.
This is an example of the approximate EM criterion discussed in
Section~\ref{sec:rate:em-da}.

\begin{figure}

\includegraphics{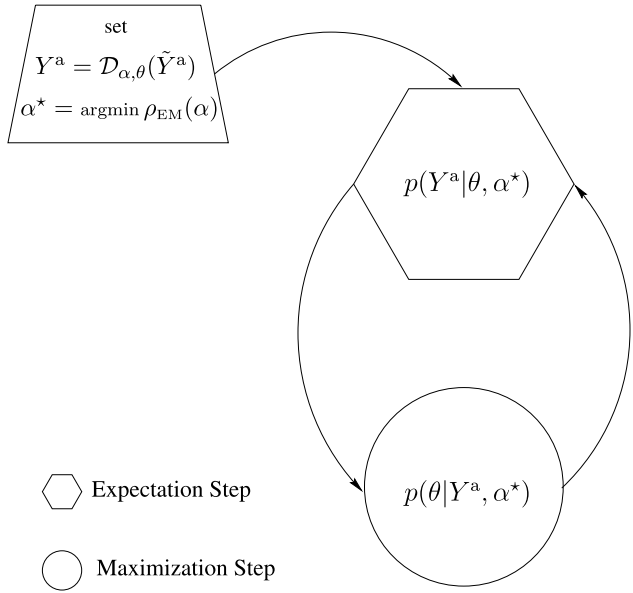}

\caption{The~EM algorithm constructed with conditional data
augmentation. [In the maximization step we compute $\theta$ to
maximize the conditional expectation of $\log p(\theta|Y^{\mathrm{aug}},\alpha^\star)$,
with the expectation computed in the expectation step. Here
we use the superscript ``a'' as an abbreviation for ``aug'' or
``augmented.'']} \label{fig:condaug}
\end{figure}

It is worth noting that the optimization required by conditional
augmentation occurs as part of the derivation of the algorithm. The~value
$\alpha$ is fixed when we run the algorithm; see Figure~\ref
{fig:condaug}. The~methods of marginal and joint augmentation, on the
other hand, avoid this initial optimization problem by averaging over
or fitting $\alpha$ on the fly, and, more importantly, they can lead
to better algorithms.

\subsection{Marginal Augmentation}
\label{sec:eda:mar}

Marginal augmentation also begins with (\ref{eq:defwkpar}), but, in
addition to a working parameter, introduces a \textit{working prior
distribution}, $p(\alpha)$. The~working prior distribution is
typically chosen so that $\alpha$ and $\theta$ are independent, so that
\vadjust{\eject}
\begin{eqnarray}\label{eq:defmaraug}
\qquad &&\int_{\mathcal{M}(Y^{\mathrm{aug}})=Y^{\mathrm{obs}}} \biggl[\int
p(Y^{\mathrm{aug}}|\theta, \alpha) p(d
\alpha)\biggr]\mu(d Y^{\mathrm{aug}})
\nonumber\\[-8pt]\\[-8pt]
\qquad &&\quad =p(Y^{\mathrm{obs}}|\theta).
\nonumber
\end{eqnarray}
Note that if we define the resulting augmented data model as
$p(Y^{\mathrm{aug}}
|\theta) = \int p(Y^{\mathrm{aug}}|\theta, \alpha) p(d \alpha)$,
we obtain
$\int p(Y^{\mathrm{aug}}|\theta)\mu(d Y^{\mathrm{aug}})=p(Y^{\mathrm{obs}}|\theta)$.
Thus, (\ref{eq:defmaraug}) results in a legitimate data augmentation scheme.
[Marginal augmentation was introduced by \citet{mengvand99} and is
very closely related to the PX-DA sampler of \citet{liuwu99}.]

This strategy is motivated by a desire to reduce the information in
$Y^{\mathrm{aug}}$ for $\theta$. Since conditioning tends to increase
information, marginalization may be advantageous. In particular, for
any function $h(\cdot)$, we have
%
\begin{eqnarray}\label{eq:marmath}
&&\mathrm{E}[\operatorname{Var}(h(\theta )|Y^{\mathrm
{aug}})|Y^{\mathrm{obs}}]\nonumber\\
&&\quad = \mathrm{E}[\mathrm{E}[\operatorname{Var}(h(\theta )|Y^{\mathrm
{aug}},\alpha)|Y^{\mathrm{obs}},\alpha]|Y^{\mathrm{obs}}]\\
&&\qquad {}+ \mathrm{E}[\operatorname{Var}[\mathrm{E}(h(\theta )|Y^{\mathrm
{aug}},\alpha)|Y^{\mathrm{aug}}]|Y^{\mathrm{obs}}]\nonumber.
\end{eqnarray}
If $p(Y^{\mathrm{aug}}|\theta,\alpha)$ is generated by
$Y^{\mathrm{aug}}=\mathcal{D}_{\alpha}({\tilde Y^{\mathrm{aug}}})$ using
the baseline augmentation,
${\tilde Y^{\mathrm{aug}}}$, then
$\mathrm{E}[\operatorname{Var}(h(\theta)|\break Y^{\mathrm{aug}},\alpha)|
Y^{\mathrm{obs}},\alpha]$ does not depend on $\alpha$ and
(\ref{eq:marmath}) implies
\begin{eqnarray*}
&&\mathrm{E}[\operatorname{Var}(h(\theta)|Y^{\mathrm
{aug}})|Y^{\mathrm{obs}}]\\
&&\quad  \geq\mathrm{E}[\operatorname
{Var}(h(\theta)
|Y^{\mathrm{aug}},\alpha)|Y^{\mathrm{obs}},\alpha]
\end{eqnarray*}
for any $\alpha$, and, thus, in terms of the geometric rate, marginal
augmentation is superior to conditional augmentation (\citeauthor{mengvand99},
\citeyear{mengvand99}). This result, however, depends on the working parameter
being introduced via $Y^{\mathrm{aug}}=\mathcal{D}_{\alpha}({\tilde Y^{\mathrm{aug}}})$,
a transformation depending
only on $\alpha$. When the transformation depends on the model
parameters as well, conditional augmentation can be superior. See
\citet{mengvand99} or \citet{liuwu99} for details.

Although there is no need to choose $\alpha$ when using marginal
augmentation, we are left with the choice of working prior
distributions. One strategy for choo\-sing $p(\alpha)$ (\citeauthor{vandmeng01art},
\citeyear{vandmeng01art}) suggests parameterizing the working prior, $p(\alpha|\psi)$,
and chooses $\psi$ as a level-two working parameter via a
conditional augmentation criterion. \citet{liuwu99} show that, under
certain conditions, the Haar measure leads to an optimal algorithm with
the correct stationary distribution. In general, however, using an
improper working prior distribution may not even lead to the correct
stationary distribution, let alone optimality; see \citet{mengvand99},
\citet{vandmeng01art} and \citet{vand09}. When it exists, the
use of the Haar measure typically leads to a joint chain on the
enlarged space $(\alpha, \theta, Y^{\mathrm{aug}})$ that is nonpositive
recurrent, but the marginal chain on the original space $\theta$ converges
properly to the desired posterior distribution
$p(\theta|Y^{\mathrm{obs}})$; see \citet{hobe01}, \citet{marchobe04} and
\citet{hobemarc08} for additional discussion.

\subsection{Joint Augmentation}
\label{sec:eda:joint}

\begin{figure}

\includegraphics{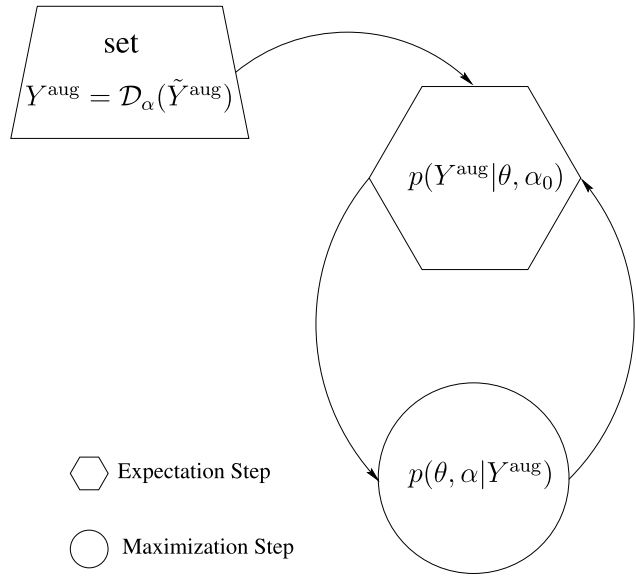}

\caption{The~PXEM Algorithm. [In the maximization step we compute
$\theta$ and $\alpha$ to maximize the conditional expectation of
$\log p(\theta,\alpha|Y^{\mathrm{aug}})$, with the expectation
computed in the
expectation step.]} \label{fig:pxem}
\end{figure}

There is no known easy way to implement EM-type algorithms that use
marginal augmentation. A~similar strategy, however, uses the
augmentation scheme (\ref{eq:defwkpar}), but rather than optimizing
$\rho_{\mathrm{EM}}$ as a function of $\alpha$ before running the
algorithm or margi\-nalizing $\alpha$ out as in (\ref{eq:defmaraug}),
this method fits $\alpha$ jointly with $\theta$ in the M-step. In
particular, \citet{liurubiwu98} presents the PXEM algorithm as a
fast adaptation of conditional augmentation in the context of the EM
algorithm in the case when $p(\theta )\propto1$, for example, in maximum
likelihood estimation. Van Dyk (\citeyear{vand00px}) slightly
extended the framework to the Bayesian case, by defining
\begin{eqnarray*}
&&Q_{\mathrm
{px}}(\theta
,\alpha|\theta',\alpha_0)\\
&&\quad = \int\log[p(Y^{\mathrm{aug}}|\theta
,\alpha
)p(\theta)]\\
&&\qquad \hphantom{\int}
{}\cdot p(Y^{\mathrm{aug}}| Y^{\mathrm{obs}},\theta',\alpha
_0)\,dY^{\mathrm{aug}}.
\end{eqnarray*}
As
illustrated in Figure~\ref{fig:pxem}, the PXEM iteration sets $(\theta
^{(t+1)},\alpha^{(t+1)})$ equal to the maximizer of
$Q_{\mathrm{px}}(\theta,\break \alpha|\theta^{(t)},\alpha_0)$,
where $\alpha_0$ is some fixed value.%
\footnote{We need not condition on $\alpha=\alpha^{(t)}$ in
$Q_{\mathrm{px}}$ because\break
$Q_{\mathrm{px}}(\theta,\alpha|\theta',\alpha') \geq
Q_{\mathrm{px}}(\theta',\alpha'|\theta',\alpha')$
implies $p(\theta|Y^{\mathrm{obs}})\geq p(\theta'|Y^{\mathrm{obs}})$
for any values of $\theta'$ and $\alpha'$. In particular,
$Q_{\mathrm{px}}(\theta^{(t+1)},\alpha^{(t+1)}| \theta^{(t)},\alpha_0)\geq
Q_{\mathrm{px}}(\theta^{(t)},\alpha_0 | \theta^{(t)}, \alpha_0)$
implies $p(\theta^{(t+1)}| Y^{\mathrm{obs}})\geq p(\theta^{(t)}| Y^{\mathrm{obs}})$;
see \citet{liurubiwu98} and \citet{vand00px}.
}
The~\textit{particular} value
of $\alpha_0$ is generally irrelevant for a PXEM iteration and is
simply set to some convenient value throughout the iteration (e.g.,
$\alpha_0=1$ for scale working parameters and $\alpha_0=0$ for
location working parameters). In this regard, the PXEM iteration could
be rewritten to avoid the dependence on~$\alpha_0$, but it is
generally deemed easier to simply set $\alpha_0$ at one arbitrary
value and avoid potentially complex algebraic manipulations. The~situation is similar when using marginal augmentation with an improper
working prior distribution. In that case the posterior distribution of
$\alpha$ is improper leading to the technical concerns discussed in
Section~\ref{sec:eda:mar}. With PXEM the observed data likelihood does
not depend on $\alpha$ which can lead to numerical problems if the
updated value of $\alpha$ is carried forward in the iteration.

We expect PXEM to perform at least as well as an algorithm that fixes
$\alpha$ (i.e., CDA-EM) in terms of the global rate of convergence
because it essentially removes the conditioning on $\alpha$ in the
data-augmentation scheme. Removing this conditioning reduces
$I^{\mathrm{aug}}$
(in a positive semidefinite ordering sen\-se) and thus improves the rate
of convergence of EM (see Meng and van Dyk, \citeyear{mengvand97}, and Liu, Rubin and
Wu, \citeyear{liurubiwu98}, for details). It is in this regard that PXEM is an example of
efficient data augmentation: it effectively reduces the augmented data
information in order to improve the rate of convergence without
sacrificing simplicity or stability. This does not mean that PXEM
generally dominates a CDA-EM algorithm because different augmentation
schemes are used in the context of the two strategies. In particular,
like marginal data augmentation, PXEM is generally implemented with a
transformation,
$Y^{\mathrm{aug}}=\mathcal{D}_{\alpha}({\tilde Y^{\mathrm{aug}}})$.
However, unlike that of conditional
data augmentation, this transformation does not depend on $\theta$;
see Figure~\ref{fig:pxem}. Liu, Rubin and Wu (\citeyear{liurubiwu98}) give an
alternative explanation for the efficient performance of PXEM, that by
fitting $\alpha$, we are performing a covariance adjustment to
capitalize on information in the data-augmentation scheme. They also
illustrate the substantial computational advantage PXEM can offer over
other EM-type algorithms for ML estimation. In the context of Bayesian
calculations, \citet{vandtang03} show how one-step-late methods
(\citeauthor{gree90}, \citeyear{gree90}) can be used to accomplish the required optimizations
of the PXEM M-step.

\subsection{A~Graphical Comparison of CDA-EM and PXEM}
\label{sec:eda:comp}

\begin{figure*}[t]

\includegraphics{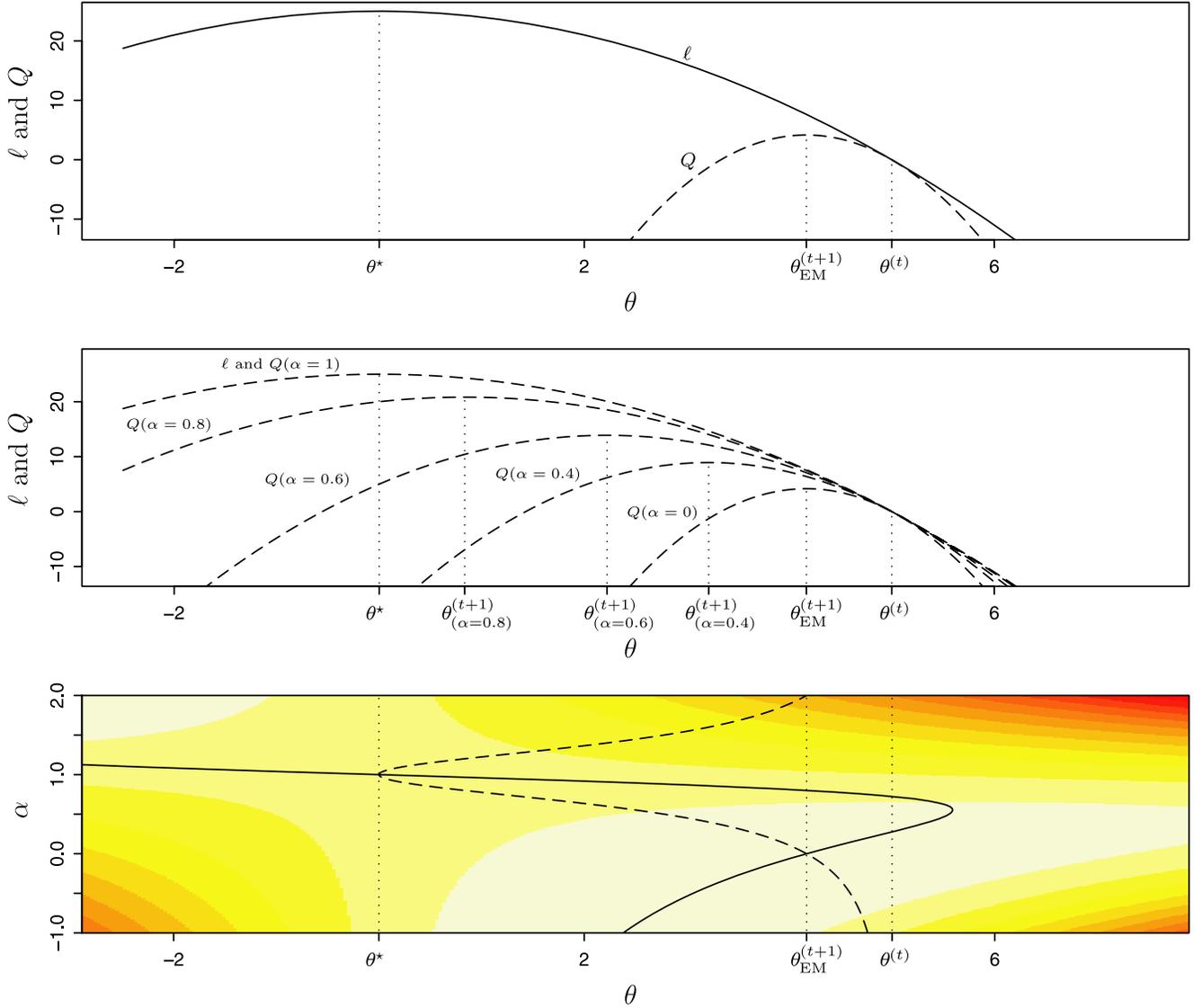}

\caption{Comparing the CDA-EM and the PXEM algorithms. The~first panel
compares $\ell(\theta)$ and $Q(\theta|\theta^\prime)$ for an EM
algorithm applied to a simple Gaussian problem. The~functions are
normalized to be tangent at $\theta^{(t)}$. The~second panel compares
$\ell(\theta)$ with $Q(\theta,\alpha| \theta^{(t)},\alpha)$ for
several values of $\alpha$ and shows how the missing data becomes less
informative and $Q(\theta,\alpha| \theta^{(t)},\alpha)$ becomes a
better approximation to $\ell(\theta)$ as $\alpha$ get closer to the
optimal value, $\alpha=1$. The~final plot is a heat map of
$Q(\theta,\alpha| \theta^{(t)},\alpha^\prime=0)$ that is optimized in the
PXEM algorithm. Lighter colors correspond to higher functional values.
The~function has two critical points, one at $(\theta=0, \alpha=1)$
and one at $(\theta=0, \alpha=-\infty)$. The~solid and dashed curves
give the optimal value of $\theta$ as a function of $\alpha$ by
maximizing $Q(\theta,\alpha| \theta^{(t)},\alpha)$ and $Q(\theta
,\alpha| \theta^{(t)},\alpha^\prime=0)$, respectively. The~CDA-EM
update is a saddle point of
$Q(\theta,\alpha| \theta^{(t)},\alpha^\prime=0)$ and the PXEM update
occurs in the limit as $\alpha\to-\infty$. Nonetheless,
both algorithms return the maximum likelihood
estimate in one iteration, $\theta^{(t+1)}=0$.}\vspace*{-6pt}
\label{fig:comp}
\end{figure*}

To illustrate the differences between the CDA-EM and PXEM algorithms,
we consider a simple Gaussian model. Suppose
%
\begin{equation}
X_i \sim\mathrm{N}(\theta, 1/2)\quad  \mbox{for } i=1,\ldots, n
\end{equation}
and
%
\begin{equation}
Y_i \sim\mathrm{N}(\theta, 1/2)\quad  \mbox{for } i=1,\ldots, m,
\end{equation}
where the $X=(X_1,\ldots, X_n)$ is observed and $Y=(Y_1, \ldots, Y_m)$
is completely missing. Obviously, the maximum likelihood estimate of
$\theta$ is $\bar X$ and the missing $Y$ is not relevant. Nonetheless,
for illustration, we can construct an EM algorithm that treats $Y$ as
missing data. In particular, with $(X,Y)$ being the augmented data, we have
\begin{eqnarray*}
Q\bigl(\theta| \theta^{(t)}\bigr) &=& 2\theta\Biggl[n\bar X + \sum_{i=1}^m\mathrm{E}\bigl(Y_i|\theta^{(t)}\bigr)\Biggr] - (n+m)\theta^2\\
&=& 2\theta\bigl(n\bar X + m\theta
^{(t)}\bigr) - (n+m)\theta^2,
\end{eqnarray*}
which can be compared to the observed data loglikelihood, $\ell(\theta
)$, as in the first panel of Figure~\ref{fig:comp}, where $n=1$,
$m=5$, $\bar X=0$, and $\theta^{(t)}=5$. The~panel illustrates that
$\ell(\theta)$ and $Q(\theta|\theta^{(t)})$ have the same derivative
at $\theta^{(t)}$ and that their optimizers are the maximum likelihood
estimate, $\theta^\star$, and $\theta^{(t+1)}_{\mathrm{EM}}$, respectively.
(For diagrams illustrating EM's iteration and rate of converge, see
\citeauthor{nav97}, \citeyear{nav97}.)

To use CDA-EM and PXEM, we introduce a working parameter $\alpha$, via
the transformation,
$Z_i = Y_i - \alpha\theta\sim\mathrm{N}[(1-\alpha)\theta, 1/2]$ for $i=1,\ldots, m$,
and treat $Z=(Z_1,\ldots, Z_m)$
as the missing data.\vadjust{\goodbreak} Since $\alpha$ is not identifiable given $X$,
it is a valid working parameter. In this case,\vspace*{-1pt}
\begin{eqnarray*}
Q\bigl(\theta,\alpha| \theta^{(t)},\alpha^\prime\bigr)
&=& 2\theta\Biggl[n\bar X + (1-\alpha)\sum_{i=1}^m\mathrm{E}\bigl(Z_i|\theta
^{(t)},\alpha^\prime\bigr)\Biggr]\\
&&{} - [n + m(1-\alpha)^2]\theta^2 \\
&=& 2\theta\bigl[n\bar X + m(1-\alpha)(1-\alpha^\prime)\theta^{(t)}
\bigr]\\
&&{} - [n + m(1-\alpha)^2]\theta^2.\vadjust{\goodbreak}
\end{eqnarray*}
The~method of conditional data augmentation requires
$I^{\mathrm{aug}}(\alpha)=2[n+m(1-\alpha)^2]$ be computed by differentiating
$Q(\theta,\alpha| \theta^{(t)},\alpha^\prime)$ twice with respect to $\theta$
and minimized it as a function of $\alpha$. The~optimal value occurs
when $\alpha=1$, in which case the distribution of the missing data
does not depend on $\theta$. The~second panel of Figure~\ref{fig:comp} compares
$Q_\alpha(\theta|\theta^{(t)})\equiv Q(\theta,\alpha| \theta^{(t)},\alpha)$
computed with several values of $\alpha$ with $\ell(\theta)$.
As $\alpha$ grows closer to one, $\theta^{(t+1)}$ grows closer to
$\theta_{\mathrm{MLE}}$. With the optimal
value of $\alpha$ in this example, $Q_\alpha(\theta|\theta^{(t)})$
and $\ell(\theta)$ coincide, and CDA-EM converges to $\theta^\star$
in one iteration. In general, the algorithm does not converge in one
step, but the underlying strategy of choosing a working parameter so
that $Q_\alpha(\theta|\theta^{(t)})$ is closer to $\ell(\theta)$ is
always the goal.

For PXEM, $\alpha^\prime$ is fixed at the identity value of the
transformation from $Y$ to $Z$ (i.e., $\alpha^\prime=0$) and $\theta$
and $\alpha$ are updated at each iteration by jointly optimizing
$Q(\theta,\alpha| \theta^{(t)},\alpha^\prime=0)$. The~third panel
of Figure~\ref{fig:comp} plots this function using a heat map, where
brighter colors represent higher values and darker colors represent
lower values. The~solid line superimposed on the plot is the optimal
value of $\theta$ as a function of $\alpha$ and is given by
%
\begin{equation}
{\sum_{i=1}^n X_i + m(1-\alpha) \theta^{(t)}\over n + m(1-\alpha)^2}.
\end{equation}
For example, with $\alpha=0$ the curve gives $\theta^{(t+1)}_{\mathrm{EM}}$.
The dashed line gives the optimal value of $\theta$ as a function of
$\alpha$ under CDA-EM. This curve corresponds to the modes of the
dashed curves in the second panel. The~solid and dashed curves in the
third panel differ because CDA-EM and PXEM differ in how they treat
$\alpha^\prime$ in $Q(\theta,\alpha| \theta^{(t)},\alpha^\prime)$.
PXEM fixes $\alpha^\prime$ at the identity value under the
transformation from $Y$ to $Z$ (i.e., PXEM fixes $\alpha^\prime=0$),
whereas CDA-EM does not update $\alpha$ in the iteration and sets
$\alpha^\prime=\alpha$ throughout. The~function
$Q(\theta,\alpha|\theta^{(t)},\alpha^\prime=0)$ plotted in panel 3 increases
along the solid curve as $\alpha$ goes to $-\infty$ and the solid curve
asymptotes to $\theta=0$, the maximum likelihood estimate. Thus, both
CDA-EM run with $\alpha=1$ and PXEM converge to the maximum likelihood
estimate in one iteration.

One might be tempted to think that PXEM is superior to CDA-EM because
it optimizes $Q(\theta,\alpha| \theta^{(t)},\break \alpha^\prime=0)$ over
both $\theta$ and $\alpha$ at each iteration, whereas CDA-EM
optimizes $Q(\theta,\alpha| \theta^{(t)},\alpha^\prime=\alpha)$
over only $\theta$ under a constraint that fixes $\alpha$ at a
prespecified value. That is, one might expect PXEM to increase $\ell$
more because it increases $Q$ more. This reasoning, however, not only
blurs the difference in how the two algorithms treat $\alpha^\prime$,
but also oversimplifies the rates of convergence of EM-type algorithms.
An algorithm that increases $Q$ more at every iteration does not
necessary converge faster. This can be seen clearly in the first panel
of Figure~\ref{fig:comp}. The~optimal update is $\theta^\star$, but
$\theta^\star$ is far from the maximizer of $Q$. Our goal is not to
increase $Q$ more, but to make $Q$ a better approximation of the log
likelihood. As another example, the EM algorithm by definition
increases $Q$ by at least as much in its \mbox{M-step} as ECM can in a
sequence of CM-steps. Nonetheless, \citet{meng94b} shows that ECM
can converge faster than EM. In the present example, CDA-EM sets
$\alpha=1$ and updates $\theta$ to $\theta^{(t+1)}=0$ which is a saddle
point of $Q(\theta,\alpha| \theta^{(t)},\alpha^\prime=0)$. Even
though $Q(\theta,\alpha| \theta^{(t)},\alpha^\prime=0)$ evaluated
at the CDA-EM update is less than when it is evaluated at the PXEM
update, both updates have $\theta^{(t+1)}=0$ and thus give the same value
of the observed data log likelihood. The~rate of convergence is more
directly determined by (\ref{eq:emrate}) than by the relative increase
in $Q$. It is this rate that CDA-EM aims to optimize and that PXEM
improves by eliminating the conditioning on $\alpha$; see Section~\ref
{sec:eda:joint}.

\begin{figure}[b]

\includegraphics{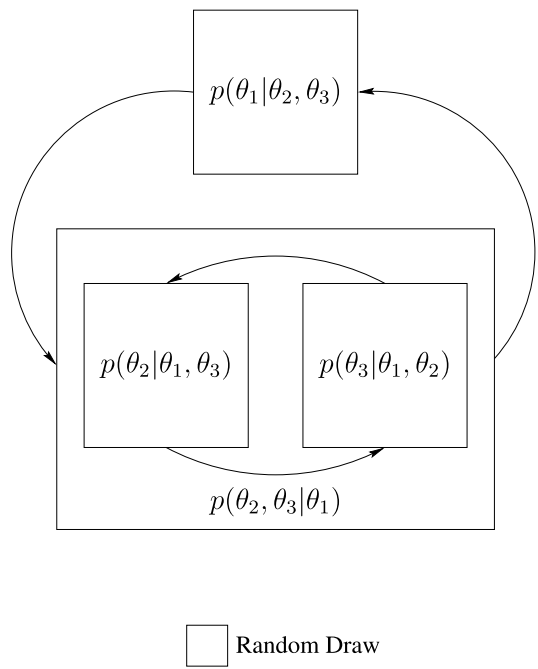}

\caption{The~partially blocked Gibbs sampler. The~inner loop is
iterated $N$ times.}
\label{fig:partblockgibbs}
\end{figure}

\subsection{Nesting}
\label{sec:eda:nest}

\begin{figure}

\includegraphics{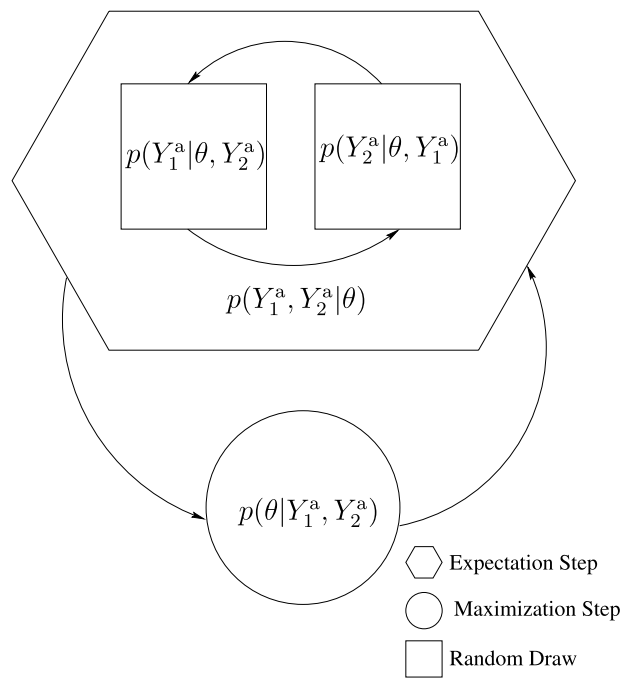}

\caption{The~MCEM algorithm.
The~inner loop is iterated several times. [In the maximization
step we compute $\theta$ to maximize the conditional expectation of
$\log p(\theta|Y^{\mathrm{aug}}_1, Y^{\mathrm{aug}}_2)$, with the
expectation computed in
the Monte Carlo expectation step. Here we use the
superscript ``a'' as an abbreviation for ``aug'' or ``augmented.'']} \label{fig:mcem}
\end{figure}

\begin{figure}

\includegraphics{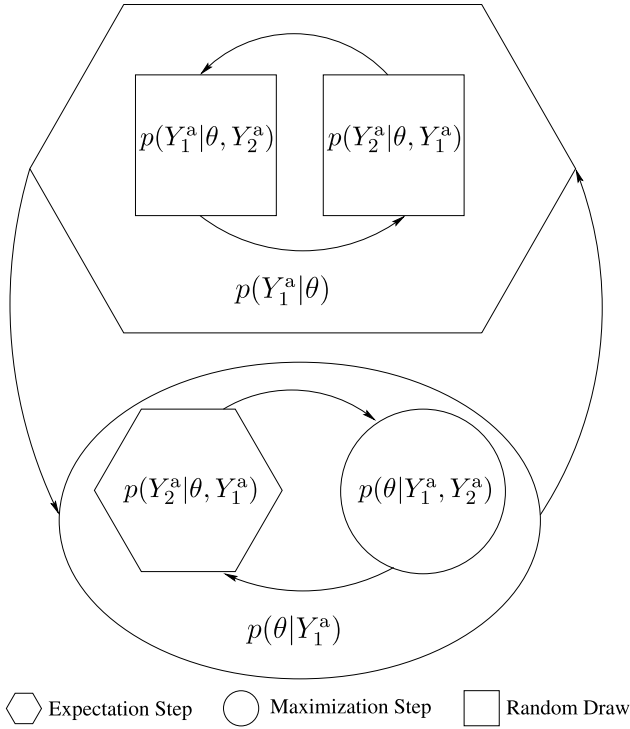}

\caption{The~nested EM algorithm with Monte-Carlo E-step
implemented with a two-step Gibbs sampler. The~inner loops are both iterated several times. [In the maximization step we
compute $\theta$ to maximize the conditional expectation of
$\log p(\theta|\break Y^{\mathrm{aug}}_1, Y^{\mathrm{aug}}_2)$, with the
expectation computed in the
expectation steps, see Section~\protect\ref{sec:eda:nest}. Here we use the
superscript ``a'' as an abbreviation for ``aug'' or
``augmented.'']}
\label{fig:nem}
\end{figure}

Nested EM and DA-type algorithms involve iteratively using a data
augmentation method to accomplish one of the steps of a larger
algorithm also involving data augmentation. Figures~\ref
{fig:partblockgibbs}--\ref{fig:nem} illustrate three different ways
this might be done. To motivate the nesting strategy, we begin with the
partially-blocked Gibbs sampler illustrated in Figure~\ref{fig:partblockgibbs}
(\citet{vand00nem}). Although we consider a
sampler composed using three full conditional distributions, the ideas
apply immediately to samplers with arbitrarily many conditional
distributions. In particular, suppose we wish to sample from
$p(\theta|Y^{\mathrm{obs}})$, where $\theta =(\theta _1,\theta _2,\theta_3)$
by using a Gibbs sampler which samples from each of
$p(\theta _1|\theta _2,\theta_3,Y^{\mathrm{obs}})$,
$p(\theta_2|\theta _1,\theta _3,Y^{\mathrm{obs}})$, and
$p(\theta _3|\theta_1,\theta _2, Y^{\mathrm{obs}})$ in
turn. If sampling from
$p(\theta _1|\theta _2,\theta _3,\break Y^{\mathrm{obs}})$ is expensive
relative to sampling from the other two conditional distributions, it
may be beneficial to sample once from
$p(\theta _1|\theta _2,\theta_3,Y^{\mathrm{obs}})$
and then to sample from $p(\theta _2|\theta _1,\theta _3,Y^{\mathrm{obs}})$
and $p(\theta_3|\theta _1,\theta _2,Y^{\mathrm{obs}})$ $N$ times each in turn. If
$N$ is large, the
internal Gibbs sampler delivers an approximate draw from the joint
distribution $p(\theta _2,\theta _3 | \theta _1)$. If this
approximation is
good, we are essentially running a blocked Gibbs sampler with
conditional distributions
$p(\theta _1|\theta _2,\break\theta_3,Y^{\mathrm{obs}})$ and
$p(\theta_2,\theta _3|\theta _1,Y^{\mathrm{obs}})$. The~partially blocked
Gibbs sampler is
useful when the advantage of blocking outweighs the cost of sampling
from $p(\theta _2,\theta _3|\theta _1,\break Y^{\mathrm{obs}})$ via a
nested Gibbs sampler. This
strategy may be helpful when $\theta _2$ and $\theta _3$ exhibit significant
correlation given $\theta _1$ and/or
$p(\theta _1|\theta _2,\theta_3, Y^{\mathrm{obs}})$ is
particularly difficult to sample (e.g., \citeauthor{vandconnkashsiem01},
\citeyear{vandconnkashsiem01}). Notice there is a subtle tradeoff
here. If $\theta_2$ and $\theta_3$ are (nearly) conditionally
independent given $\theta_1$, then there is no need to run the inner
iteration. If, on the other hand, they are highly correlated, then the
inner iteration may need to be run many times in order to deliver a
good draw. The~key to success with this strategy is repeating the
expensive draw of $p(\theta_1 | \theta_2, \theta_3)$ as seldom as possible.

In the context of the EM algorithm, we can implement a similar strategy
when the augmented data naturally divide into two or more parts. This
strategy takes advantage of the fact that an EM algorithm that treats
only part of $Y^{\mathrm{aug}}$ as missing and collapses over the rest
is faster
in terms of $\rho_{\mathrm{EM}}$ (\citeauthor{mengvand97}, \citeyear{mengvand97}). Thus, we aim to
construct an EM algorithm using only part of $Y^{\mathrm{aug}}$.
Although this
algorithm typically does not have a closed form M-step, the
maximization can be accomplished by a second, typically closed-form, EM
algorithm that treats the remainder of $Y^{\mathrm{aug}}$ as missing
data. The~resulting nested EM algorithm (\citeauthor{vand00nem},
\citeyear{vand00nem}) has an improved rate
of convergence but, because of the nesting, each iteration requires
more time to compute. If the computational complexity of the \mbox{E-step} is
relegated to the outer loop, this trade-off can go in favor of the
nesting strategy when considering the actual computing time required.
This advantage can be pronounced when the outer E-step requires a Gibbs
sampler to compute the necessary conditional expectations. This is
possible with the Monte Carol EM (MCEM) algorithm (\citet{weitann90}),
as is illustrated by \citet{vand00nem}. The MCEM algorithm is compared
with the nested EM algorithm in Figures~\ref{fig:mcem} and \ref{fig:nem}.

\section{Partial Collapsing as a~Unified~Approach}
\label{sec:pcm}

While the partially-blocked nature of the sampler in Figure~\ref
{fig:partblockgibbs} is clear, the nested EM algorithm in Figure~\ref
{fig:nem} partially removes
$\tilde Y^{\mathrm{aug}}\setminus Y^{\mathrm{aug}}_1 \subset Y^{\mathrm{aug}}_2$
from the data augmentation scheme in the spirit of
conditional augmentation. In this regard, the nested EM algorithm is a
type of ``partially collapsed'' EM algorithm. In this section we
discuss a different strategy for partially collapsing quantities out of
an EM or DA algorithm. In particular, in algorithms that involve model
reduction, we can collapse quantities in some but not all of the
CM-steps or conditional draws. It is in this sense that we use the
term ``partially collapsed.''

Collapsing involves constructing an algorithm on a marginal
distribution of the target space of the original algorithm. That is, we
construct an algorithm that works on a \textit{collapsed parameter space}
of the \textit{original parameter space}. (Here the parameter space
includes all unknowns including latent variables and missing data.)
Although this strategy is computational efficient it can be practically
difficult if some or all of the full conditional distributions on the
collapsed parameter space are complex or nonstandard distributions.
Given that the augmented data are introduced specifically to simplify the
full conditional distributions, it is not surprising that reducing that
augmented data can sacrifice this simplicity. Partially collapsed
methods aim to reap some of the gains of collapsing in this situation.
In particular, when some of the conditional distributions on the
collapsed parameter space are simple or at least no more complicated
that the corresponding conditional distribution of the original
parameter space, partially collapsed methods mix conditional
distributions from the two (or perhaps more) parameter spaces in the
construction of EM-type algorithms and DA-type samplers. For example,
if a conditional maximization or draw given the augmented data are not
easier than the corresponding maximization or draw given the observed
data, then we may as well use the version that does not involve data
augmentation, that is the collapsed version. As we shall discuss, this
strategy has lead to a number of useful algorithms.

\subsection{The~ECME and AECM Algorithms}
\label{sec:pcm:aecm}

In order to improve the rate of convergence of the ECM algorithm,
\citet{liurubi95} formulated the Expectation Conditional
Maximization Either or ECME algorithm in which they suggest replacing
one or more of the CM-steps of the ECM algorithm with
\begin{enumerate}
\item[Direct CM-step $p$:] Set
$\theta^{(t+{p/ P})} = \operatorname{argmax}_\theta \log p(\theta|\break Y^{\mathrm{obs}})$
\end{enumerate}
subject to $\theta_{-p}^{(t+{p/ P})} = \theta_{-p}^{(t+{(p-1)/ P})}$.
When an iterative method is required to accomplish one or more of the
CM-step of ECM, it is often no more difficult to maximize the
conditional log posterior directly without recourse to data
augmentation. In this case \citet{liurubi95} argue that the direct
\mbox{CM-step} is expected to improve convergence without complicating
implementation. We recognize this as a partially collapsed algorithm.
If all of the ECM \mbox{CM-steps} were replaced by direct \mbox{CM-steps} the
augmented data would be completely removed from the iteration. This
would collapse ECM into a Gauss--Seidel optimizer, which is generally
expected to be faster than ECM. Of course, if some of the CM-steps of
ECM are simple closed-form optimizations while those of ECME require
numerical optimization, the computational tradeoff can easily favor ECM
over Gauss--Seidel.

\begin{figure}

\includegraphics{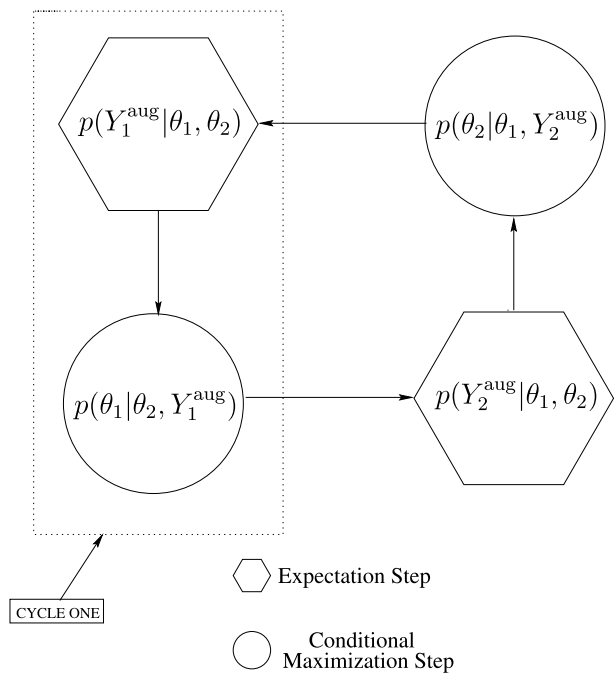}

\caption{A~two-cycle AECM algorithm. [In the conditional maximization
steps, we compute the component of $\theta=(\theta_1,\theta_2)$ to
maximize the conditional expectation of the log of the quantity in the
$\bigcirc$, with the expectation computed in the most recent
expectation step, see Section~\protect\ref{sec:pcm:aecm}.]}
\label{fig:aecm}
\end{figure}

\citet{mengvand97} set up a more general framework by allowing
different levels of augmented data in each CM-step. The~resulting
algorithm is called the Alternating Expectation Conditional
Maximization or AECM algorithm and generalizes both the ECME and the
SAGE (\citet{fesshero94}) algorithms. In particular,
Meng and van Dyk suggest replacing the CM-step of ECM with
\begin{enumerate}
\item[CM-step $p$:] Set
$\theta^{(t+{p/ P})} = \operatorname{argmax}_\theta\mathrm{E}
[\log p(\theta | \break g_p(Y^{\mathrm{aug}}) ) | \theta^{(t+{(p-1)/P})} ]$
\end{enumerate}
subject to $\theta_{-p}^{(t+{p/P})} = \theta_{-p}^{(t+{(p-1)/P})}$.
Here we have expanded $Q(\theta| \theta^{(t)})$ according to
its original definition with two important changes. First, $Y^{\mathrm{aug}}$ is
replaced by some function $g_p$ of $Y^{\mathrm{aug}}$. This allows us
to reduce
the data augmentation by differing amounts in each of the $P$ CM-steps.
Here we assume $g_p(Y^{\mathrm{aug}})$ is a legitimate data
augmentation scheme
for each $p$. In particular, $Y^{\mathrm{obs}}$ is part of each
$g_p(Y^{\mathrm{aug}})$.
Second, because the data augmentation varies among the CM-steps, we
must compute and E-step each time the data augmentation changes, see
Figure~\ref{fig:aecm}. Thus, in the expectation of each AECM CM-step
we condition on the value of $\theta$ produced by the most recent
CM-step, not the value produced at the end of the previous iteration.
If the data augmentation is the same for several consecutive \mbox{CM-steps}
(i.e., if $g_p$ is the same) we need only recompute the E-step at the
beginning of this sequence. The~same requirement holds for ECME in that
the steps must be appropriately ordered relative to the E-step. The~CM-steps
that involve data augmentation must all follow the E-step
and be performed before any of the CM-steps that do not involve data
augmentation, unless the E-step is repeated. These step-ordering
requirements are necessary to ensure monotone convergence of the ECME
and AECM algorithms (\citeauthor{mengvand97}, \citeyear{mengvand97}). As we
discuss next, similar
step-ordering requirements apply to the partially collapsed Gibbs sampler.

\subsection{The~Partially Collapsed Gibbs Sampler}
\label{sec:pcm:pcg}

Consider the two-step data augmentation sampler described in
Section~\ref{sec:comp:da}. To clarify ideas, we rewrite this sampler
with $Y^{\mathrm{aug}}$ replaced by $\psi$ and with the conditioning
on $Y^{\mathrm{obs}}$
suppressed:
\begin{longlist}
\item[Step~1:] $\psi^{(t+1)}\sim p(\psi|\theta^{(t)})$,\vspace*{1pt}
\item[Step~2:] $\theta^{(t+1)}\sim p(\theta| \psi^{(t+1)})$.
\end{longlist}
Under the standard regularity conditions, we expect that after
sufficient burn-in this sampler will effectively return correlated
draws from its stationary distribution, $p(\psi,\theta)$. In order to
speed up convergence to stationarity and reduce the correlation of the
draws, we might take a cue from ECME and AECM and attempt to partially
collapse the sampler. In particular, suppose we want to reduce the
conditioning in Step~2. A~reasonable and optimal strategy might seem
to be the following:
\begin{longlist}
\item[Step~1:] $\psi^{(t+1)}\sim p(\psi|\theta^{(t)})$,
\item[Step~2:] $\theta^{(t+1)}\sim p(\theta)$.
\end{longlist}
Clearly, $\psi^{(t+1)}$ and $\theta^{(t+1)}$ are independent and the
stationary distribution of this sampler is $p(\psi)p(\theta)$ which
is generally different than the target distribution, $p(\psi,\theta
)$. In this simple example, we need only change the order of the two
steps to regain a chain with the target distribution as its stationary
distribution. Nonetheless, three important cautionary facts regarding
partially collapsed Gibbs samplers are illustrated by this simple example.

First, the ``full conditional distributions'' of the partially
collapsed sampler may not be compatible with \textit{any} joint
distribution. In the simple example, this is illustrated by the fact
that one cannot find a joint distribution of $(\psi,\theta)$ such
that $\psi$ depends on $\theta$ but $\theta$ is independent of $\psi
$. This incompatibility means that we have left the standard Gibbs
sampler framework and that standard results as well as our intuition
may fail. Second, as with ECME and AECM, the \textit{order} of the steps
may matter. Even in this simple case, the stationary distribution of
the chain depends on the order of the steps.

Finally, the steps can \textit{sometimes} be blocked to form a standard
sampler. If we first draw $\theta$ from its marginal distribution and
then $\psi$ from its conditional distribution given $\theta$, we are
directly sampling from the joint distribution, and have thus blocked
the two steps. In fact, blocking is a special case of partially
collapsing. It is easy, however, to construct cases where partially
collapsed samplers do not correspond to any blocked version of the
ordinal sampler (\citeauthor{vandpark08}, \citeyear{vandpark08}; \citeauthor{parkvand09}, \citeyear{parkvand09}).

Given these cautionary facts, it is clear that care must be taken when
partially collapsing a Gibbs sampler. Van~Dyk and Park (\citeyear
{vandpark08}) give a prescriptive method for construction such
samplers that are guaranteed to maintain the target stationary
distribution. They also argue that like blocking, partial collapsing
improves the convergence characteristics of the chain, but not as much
as complete collapsing. This, along with the fact that blocking is a
special case of complete collapsing, unifies the blocking and
collapsing strategies. Generally, blocking is not as efficient as
collapsing because blocking is only partial collapsing.

\section{Refined Algorithms for the~Spectral~Model}
\label{sec:examp}

By far the most computationally intensive aspects of the EM and DA
algorithms for the spectral model described in Sections~\ref
{sec:comp:da-ex} and \ref{sec:comp:mr-ex} are the removal of the
background counts and the deblurring of the source counts, that is,
computing the conditional expectation of or sampling $Y_i^+$ and
$\dot Y_j^+$ for $i\in\mathcal{I}$ and $j \in\mathcal{J}$. These tasks involve
looking up values in the typically large matrix, $M$, a time-consuming
task even when sophisticated sparse-matrix techniques are implemented.
Given the computation cost of these steps and the hierarchical
structure of the data augmentation, nesting is an obvious strategy. As
an illustration, we implement a nested EM algorithm. In this algorithm
we start by setting $Y^{\mathrm{aug}}_1$ equal to $\dot Y_j^+$ for
$j\in\mathcal{J}$.
Because this augmentation is smaller than the complete
data-augmentation scheme outlined in Table~\ref{tbl:specda}, fewer
iterations of the EM algorithm are required. Because there is less
augmented data, however, the M-step is not in closed form. Thus, we
implement an inner EM algorithm to accomplish the M-step of the outer
EM algorithm. This strategy is similar to the algorithm illustrated in
Figure~\ref{fig:nem}, except the outer E-step does not require a
Gibbs sampler but is nonetheless computationally demanding. The~inner
EM iteration fixes $Y^{\mathrm{aug}}_1$ and updates only the first
three rows of
Table~\ref{tbl:specda} in the inner E-step and $\theta$ in the
M-step. If this inner EM converges slowly (e.g., there are many and/or
weak emission lines), a relatively large number of inner iterations
(e.g., 10) may substantially improve the speed of the algorithm. The~outer E-step updates all of $Y^{\mathrm{aug}}$.

\begin{figure}

\includegraphics[scale=0.98]{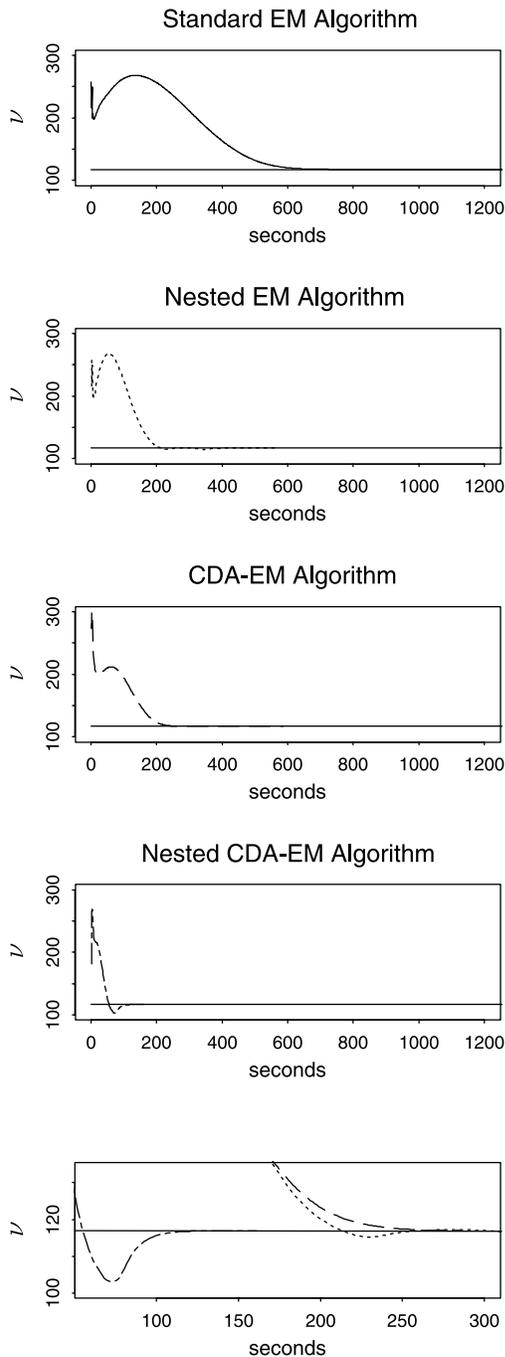}

\caption{Various EM-type algorithms for fitting the spectral model.
The~figure illustrates the computational advantage of nesting and
conditional augmentation. All five plots show the convergence of the
parameter $\nu$, the expected line count, as a function of C.P.U. time
in seconds. The~five plots correspond to the standard EM algorithm
based on the data-augmentation scheme outlined in Table~\protect\ref
{tbl:specda} (solid line); the nested EM algorithm (dotted line); the
CDA-EM algorithm (dashed line); an algorithm that combines nesting and
CDA-EM (dotted--dashed line); and a close up of the first 300 seconds
comparing all but the standard EM algorithm. The~solid horizontal line
in each plot is the MLE of $\nu$. The~nested and CDA-EM used here are
described in Section~\protect\ref{sec:examp}.}
\label{fig:examp}
\end{figure}

The~advantage of nesting is illustrated using a spectrum of the high
redshift quasar S5 $0014+81$ collected with the \textit{Chandra} \textit{X-ray
Observatory} as described by \citet{elvietal94}. The~spectrum is
modeled using a power law continuum,
$f(\theta^C,E_j)=\gamma E_j^{-\beta}$, exponential absorption,
$g(\theta^A,E_j)=e^{\xi/E_j}$, and a single Gaussian emission line with location,
width, and intensity parameters\footnote{A~Gaussian emission line is
parameterized as ${\nu\over\sigma}\phi({E-\mu\over\sigma})$, where $\phi$ is
the standard normal probability density function,
$\mu$ is the line location, $\sigma$ is the line width, and $\nu$ is
the line intensity.} for a total of six free parameters. The~first two
panels of Figure~\ref{fig:examp} show the convergence of $\nu$, the
expected counts attributed to the line, for the EM and nested EM
algorithms, respectively. The~nested EM algorithm (run with 4 inner
iterations) converges in about a third of the time required by the
standard EM algorithm. The~remaining panels in Figure~\ref{fig:examp}
will be described shortly.

To further improve the convergence of the algorithms, we can reduce the
augmented information for $\theta$ using the method of conditional
augmentation. In particular, we reduce the counts attributed to the
absorbed photons in the emission line, $\ddot Y^L_j - \dot Y^L_j$. Recall
that absorption does not occur uniformly across the range of energies
of an emission line, and the energies of the observed photons are
biased towards areas of low absorption, complicating parameter
estimation. Our typical strategy, as described in Table~\ref
{tbl:specda}, is to treat the absorbed photons as missing data. Thus,
in the augmented data, there is no absorption. It is important to note,
however, that we need not account for (i.e., augment) all of the
absorbed photons, rather we only need the absorption rate to be \textit{constant} across the support energies of the emission line. Thus, a
better strategy is to augmented fewer absorbed photons, just enough so
that the absorption rates are equal across the range of energies of an
emission line. In particular, suppose $a_{\min}$ is the lowest
absorption rate, $1-d_jg(\theta^A, E_j)$, where $j$ varies over the
support of the emission line. To reduce the volume of the augmented
data, we can compute $\ddot Y^k_j$ acting as\vspace*{-1pt} if the absorption rate were
$1-d_jg(\theta^A, E_j) - a_{\min}$. Here $a_{\min}$ is the optimal
value of a working parameter, and we condition on it throughout. In
this way, we add fewer counts to each bin. As an extreme example,
consider a delta function emission line that is contained entirely
within a single energy bin. In this case, the support of the emission
line is one bin, $a_{\min} = 1-d_jg(\theta^A, E_j)$ with $j$ the
index of the bin containing the line, $1-d_jg(\theta^A, E_j) - a_{\min
}$ is zero, and we need not impute any missing counts to account for
absorption in the line. We emphasize that this does not change the
model being fit, it only improves the efficiency of the computation.
This strategy is used in the CDA-EM algorithm and is combined with
nesting in the nested CDA-EM algorithm; both algorithms are illustrated
in Figure~\ref{fig:examp}. The~nested EM algorithm and the CDA-EM
algorithm (coincidentally) require similar computation time, combining
the two strategies, however, is twice as fast as either alone. The~final panel in Figure~\ref{fig:examp} is a more detailed comparison of
the three improved algorithms. These algorithms are discussed and
further illustrated in \citet{vandkang04}.

Other strategies described in this article lead to additional
improvements. The~posterior distribution or likelihood of the location
of a narrow emission line, for example, is typically highly multimodal.
The~Poisson nature of the data leads to small energy ranges with more
counts than expected. These correspond to possible locations of a
narrow emission line and may be relatively large modes of the
likelihood if the actual line is weak. The~standard EM and DA
algorithms described here are not able to jump between these modes
because line location is updated while conditioning on which photons
are attributed to that line. Thus, the line location will be among the
energies of these photons and only photons in this energy range will be
attributed to the line in the next step. To get around this, \citet
{vandpark04} and \citet{parkvand09} suggest EM-type and DA-type
samplers that remove the conditioning on all or part of the augmented
data \textit{while updating the line locations}. The~result is ECME and
AECM algorithms for mode finding and partially collapsed Gibbs samplers
for posterior exploration, all of which are much more efficient than
the standard EM and DA algorithms.



\section{Concluding Remarks}
\label{sec:disc}

The~highly flexible nature of multilevel modeling inhibits an
off-the-shelf algorithmic approach to model fitting. However, the
flexibility of a dynamic combination of data augmentation and model
reduction give us tools to tackle these models. As illustrated in the
spectral model, the many recent extensions and refinements of data
augmentation methods can substantially improve computational speed
while maintaining simplicity and stable convergence, thus greatly
extending the applicability and power of data augmentation .

The~data-augmentation and model-reduction strategies outlined in this
article have been used either explicitly or implicitly to derive
numerous efficient EM-type and DA-type algorithms with applications to
a wide range of models including longitudinal data analysis for binary
response and robust methods, robust regression, binary and grey-level
Ising models, dynamic linear models, finite mixture models, Poisson
image analysis, probit regression, multinomial probit models,
switching-state space models, factor analysis, spectral analysis, etc.
A~small subset of examples can be found in
\citeauthor{liurubi94} (\citeyear{liurubi94}, \citeyear{liurubi95}),
\citet{gelfsahucarl95},
Meng and van Dyk (\citeyear{mengvand97}, \citeyear{mengvand98}, \citeyear{mengvand99}),
\citet{vandtang03}, \citet{vandpark04}, \citet{higd98}, \citet{pilllind01},
\citet{liurubiwu98},
\citeauthor{vand00px} (\citeyear{vand00px}, \citeyear{vand00nem}),
\citet{liuwu99}, \citet{vandmeng01art},
\citet{foulvand00}, \citet{vandkang04},
\citeauthor{imaivand05} (\citeyear{imaivand05}, \citeyear{imavand05jss}),
\citet{gelmetal08}, \citet{popewong05} and \citet{ghosduns09}. We hope that
this over\-view paper will help to both further stimulate methodological
research and promote efficient implementation of EM-type and DA-type
algorithms in practice. In other words, to paraphrase the title, we hope
practitioners will have an easier time to climb likelihood surfaces
using EM-type algorithms and to explore posterior landscape using
DA-type samplers.

\section*{Acknowledgments}
David A. van Dyk is supported in part by  NSF Grants DMS-04-06085, SES-05-50980 and DMS-09-07522.
Xiao-Li Meng is supported in part by NSF Grants DMS-04-05953, DMS-05-05595, DMS-06-\break52743 and DMS-09-07185.

\end{document}